%% file: regularization_short_term.tex
\documentclass[review]{elsarticle}

\usepackage{lineno,hyperref}

\usepackage{siunitx}
\usepackage{booktabs}
\usepackage{makecell}
\usepackage{subfigure}
\usepackage{amsthm}
\usepackage{multirow}
\usepackage{boldline}
\usepackage{bm}
\usepackage{amsmath}

\DeclareFontFamily{OT1}{pzc}{}
\DeclareFontShape{OT1}{pzc}{m}{it}{<-> s * [1.10] pzcmi7t}{}
\DeclareMathAlphabet{\mathpzc}{OT1}{pzc}{m}{it}

\DeclareMathOperator{\vect}{vec}

\DeclareMathOperator{\diag}{\emph{diag}}

\DeclareMathOperator*{\argmin}{arg\,min}

\DeclareMathOperator{\Tr}{Tr}

\modulolinenumbers[5]

\journal{Sustainable Energy Technologies and Assessments}

\begin{document}

\begin{frontmatter}

\title{Regularization methods for the short-term forecasting of the Italian electric load}


\author[unipvaddress]{Alessandro Incremona\corref{mycorrespondingauthor}}
\cortext[mycorrespondingauthor]{Corresponding author}
\ead{alessandro.incremona01@universitadipavia.it}

\author[unipvaddress]{Giuseppe De Nicolao}
\ead{giuseppe.denicolao@unipv.it}

\address[unipvaddress]{Department of Industrial and Information Engineering, University of Pavia, Via Adolfo Ferrata 5, 27100, Pavia, Italy}

\begin{abstract}
The problem of forecasting the whole 24 profile of the Italian electric load is addressed as a multitask learning problem, whose complexity is kept under control via alternative regularization methods. In view of the quarter-hourly samplings, 96 predictors are used, each of which linearly depends on 96 regressors. The $96 \times 96$ matrix weights form a $96 \times 96$ matrix, that can be seen and displayed as a surface sampled on a square domain. Different regularization and sparsity approaches to reduce the degrees of freedom of the surface were explored, comparing the obtained forecasts with those of the Italian Transmission System Operator Terna. Besides outperforming Terna in terms of quarter-hourly mean absolute percentage error and mean absolute error, the prediction residuals turned out to be weakly correlated with Terna’s, which suggests that further improvement could ensue from forecasts aggregation. In fact, the aggregated forecasts yielded further relevant drops in terms of quarter-hourly and daily mean absolute percentage error, mean absolute error and root mean square error (up to 30\%) over the three test years considered.
\end{abstract}

\begin{keyword}
Demand forecasting \sep Energy forecasting \sep Model selection \sep Neural networks \sep Regression
\end{keyword}

\end{frontmatter}

\linenumbers

\section{Introduction}

In the energy field, the availability of accurate forecasts is crucial in order to ensure the equilibrium between electricity demand and production. 
This balance must be continuously maintained within the national power grid for guaranteeing a stable electricity supply to all the consumers and preventing dangerous instabilities that could lead to blackout phenomena, which represent a serious risk from a social and economical perspective. Moreover, with the liberalization of the energy markets, load forecasts provide crucial information to players in the energy market bidding and inaccurate predictions might cause relevant financial losses.

While long-term forecasts refer to future times ranging from some months to years ahead and are relevant for long-term tasks such as planning and maintenance schedules, short-term predictions cover a period of time ranging from few minutes to one-day ahead. Typically, it is required to predict the whole 24-hour profile of the electric load, based on historical data up to the previous day and exogenous variables such as weather forecasts.

It is beyond our scopes providing a comprehensive review of the extensive literature devoted to short-term electric load forecasting. Herein, it will suffice to observe that a variety of techniques have been investigated. For what concerns classical statistical methods, some of the most popular techniques are Exponential Smoothing (ES) \cite{christiaanse1971short}, Autoregressive model (AR), Moving Average model (MA), Autoregressive Moving Average model \cite{nbamalu1993autoregressive, chen1995arma, huang1997thresholdautoregressive} (with all their variants and extensions for including seasonal phenomena, exogenous variables and so on \cite{soares2005seasonalautoregressive, yang1995armax}), non-parametric regression \cite{charytoniuk1998nonparametricregression}, semi-parametric regression \cite{hyndman2012semiparametric, dordonnat2016semiparametric}, state space models and Kalman filter \cite{moghram1989fiveshorttermforecastingtechniques, alhamadi2004kalmanfilter, takeda2016kalmanfilter}. On the other hand, machine learning and artificial intelligence techniques have been extensively applied to the energy forecasting field in the recent years, with techniques such as Artificial Neural Networks (ANNs) \cite{kuo2018artificialneuralnetwork, ryu2017deepneuralnetwork, kong2017lstmneuralnetwork, yun2008rbf, cecati2015rbfshortterm, jalali2021novel}, fuzzy logic \cite{hassan2016fuzzy, khosravi2013load}, Support Vector Machines (SVMs) \cite{zhang2017svm, jiang2018svr}.


There are multiple factors that should be considered when developing a forecasting model: while, as already said, accuracy is one of the main aspect, the complexity of the underlying model matters as well. A model considering too many parameters and predictors requires a complex tuning procedure and it is difficult to interpret, while a simpler model has the advantage of being more robust and transparent, which allows to better understand the underlying phenomenon and simplify calibration and fault detection.

The purpose of this work is the development of a whole-day ahead forecast for the quarter-hourly electric load of Italy during normal days by relying exclusively on the loads recorded in the previous days. This can be done for the load demand time series, without resorting to exogenous data such as weather forecasts, in view of its highly correlated nature \cite{hagan1987thetimeseriesapproachtoshorttermloadforecasting, sood2010autocorrelation, yadav2010autocorrelationsom, koprinska2015autocorrelation}. The analysis is restricted to so-called `normal days', that is those days without special events such as holidays. It will be shown that, after a suitable data preprocessing, consisting of a logarithmic transformation and a 7-day differentiation, accurate one-day ahead predictions of the 24-hour profile can be achieved via a weighted linear combination of the 24-hour profile of the previous day.
In view of the quarter-hourly sampling, the load time series is decomposed into 96 time series, each representing the load during a quarter-hour of a day. Accordingly, our goal is predicting the $96$-dimensional vector of loads. The preprocessing includes a logarithmic transformation, detrending, and the computation of the 7-day differenced series. This last time series represents our target to predict. Tomorrow's 96 samples of the target are modelled as the linear combination of today's 96 samples, which corresponds to a first-order Vector Autoregressive (VAR) model structure, see \cite{kilian2017var, lutkepohl2005multipletimeseries} for extensive reviews about VAR models. The identification of the predictor weights involves the estimation of a $96\times96$ matrix. However, in view of the strong correlation between consecutive quarter-hourly load values, it is reasonable to assume that the entries of the matrix behave as samples of a smooth surface. This justifies the adoption of regularization-based machine learning techniques. In the context of sparsity and regularization techniques applied to control the complexity of VAR models, one might mention tensor decomposition \cite{wang2020tensor}, LASSO shrinkage \cite{krampe2021varsparsity}, and ridge regression \cite{ballarin2021varridge}, none of which was however applied to load forecasting.\\
In this work, alternative strategies are applied in order to reduce the degrees of freedom and the variance of the parameters of the vector predictor: besides standard ridge regularization, a bidimensional penalty on the second-order differences of the weights along both directions of the surface, a regularized Radial Basis Function, and two sparsity methods, called `Two-edges model' and `One-edge model'. The obtained forecasts are compared over different test years, using as benchmark the predictor of the national Transmission System Operator (TSO) Terna, which every day within 00:00 and 00:15, publishes the quarter-hourly prediction of the Italian load demand for the entire day.

One of the main findings of this work is that, with a proper preprocessing step and a clever usage of  regularization and sparsity techniques, our linear predictor provides accurate predictions that compare favourably with our benchmark. Indeed, the new forecaster achieves significant improvements, in terms of Mean Absolute Percentage Error (MAPE), Root Mean Squared Error (RMSE) and Mean Squared Error (MAE).
In the final part of the paper it is shown that, due to the low correlation between the residuals obtained by the proposed approaches and Terna's one, a simple aggregation allows to increase the performances even further. Depending on the test year the improvement varies around $30\%$ which appears rather remarkable.\\
The paper is organized as follows: in Section \ref{sec:Dataset_and_preprocessing} the dataset and the preprocessing phase are described, while in Section \ref{sec:Problem_statement} the problem statement is formulated. Section \ref{sec:Forecasting_methods} presents the different modelling techniques adopted and Section \ref{sec:Experimental_validation_setup} describes the experiment setups. Section \ref{sec:Forecasting_results} compares and discusses the predictive performances of the proposed techniques. Finally, Section \ref{sec:Conclusions} summarizes the main results and concludes the paper.


\section{Dataset and preprocessing}
\label{sec:Dataset_and_preprocessing}

The available data consist of: (i) a 5-year long time series of quarter-hourly Italian electric load demands (from 2015 to 2019); (ii) a 3-year long time series of quarter-hourly forecasts elaborated by the national Transmission System Operator (TSO) Terna (from 2017 to 2019) \footnote{The 2015 and 2016 forecasts were also available on the Terna Transparency Report Platform. However, their forecasting error appears significantly biased, making them unusable for benchmarking purposes.}. Both datasets were downloaded from the \cite{ternaWebsite}.

\begin{figure}
\centering
\includegraphics[width=.8\textwidth]{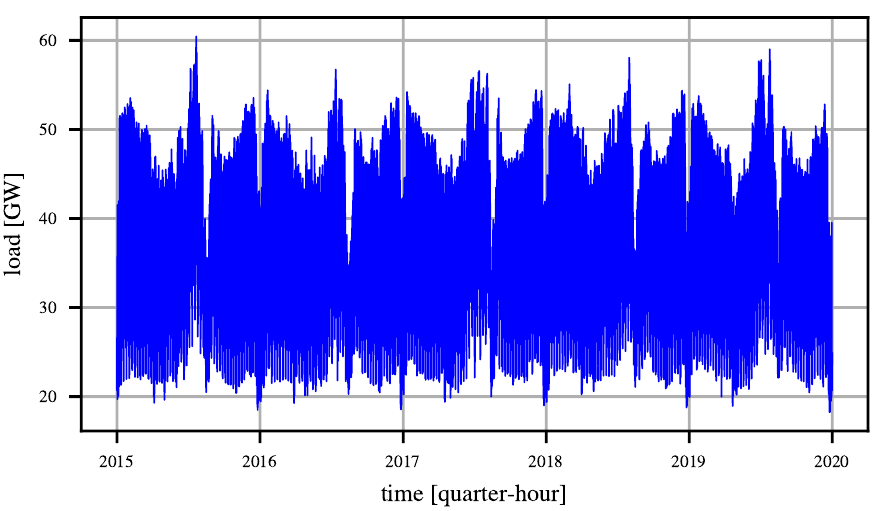}
\caption{Italian quarter-hourly electric load demand from 2015 to 2019.}
\label{fig:incre1}
\end{figure}

\begin{figure}
\centering
\includegraphics[width=.8\textwidth]{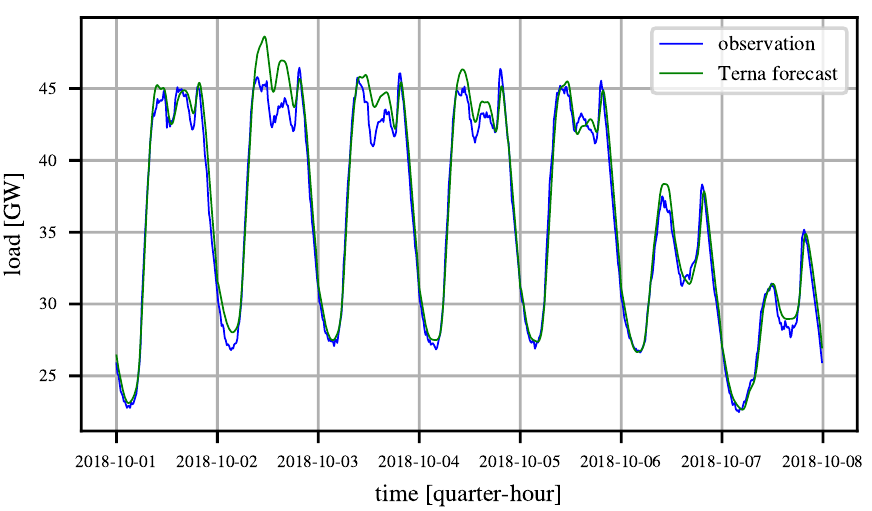}
\caption{Italian quarter-hourly electric load demand vs Terna prediction over the first week of October 2018. In spite of the good accuracy, there is room for improvement as seen on Tuesday and Wednesday, where the actual load value is underestimated.}
\label{fig:incre2}
\end{figure}

The Italian electric load demand is displayed in Fig. \ref{fig:incre1}, while Fig. \ref{fig:incre2} displays an example of one-day ahead predictions by Terna over a week. It can be seen that, although the performance is rather good, there might still be some room for improvement, which motivates the analysis of this paper.

In the following, $L(d, q)$ denotes the country load demand at the $q$-th quarter-hour of the day $d$, where $1 \le q \le 96$ and $d$ is an integer serial number representing the whole number of days from a fixed, preset date (e.g. January 0, 0000) in the proleptic ISO calendar.

In the following, when referring to the signal $L(d, q)$, we will mean the univariate time series
\begin{equation*}
\left\{\begin{array}{ccccccccc} \ldots & L(1, 1) & L(1, 2) & \ldots & L(1, 96) & L(2, 1) & \ldots & L(2, 96) & \ldots \end{array} \right\}
\end{equation*}

\begin{figure}[!ht]
\centering
  \subfigure{\includegraphics[width=.8\textwidth]{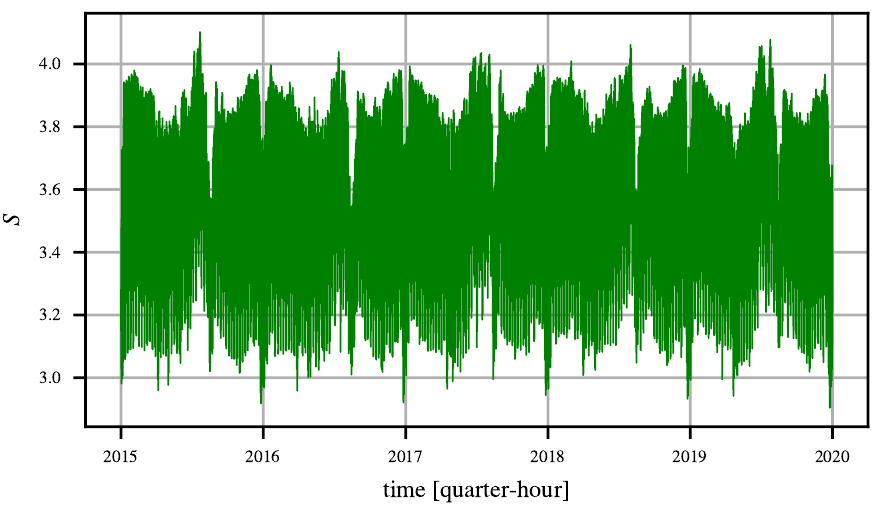}}
  \subfigure{\includegraphics[width=.8\textwidth]{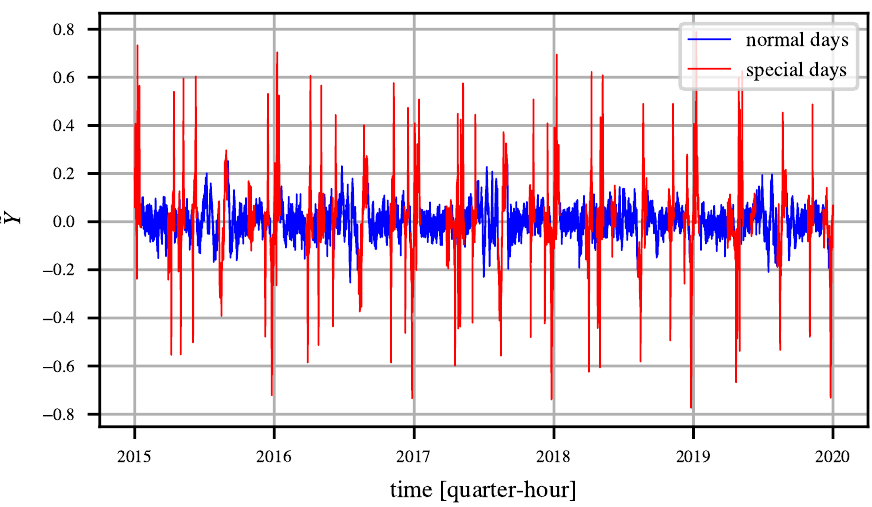}}
\caption{Preprocessed Italian quarter-hourly electric load demand:  log-transformed time series $S$ (top) and 7-day difference of log-transformed time series $\tilde{Y}$ (bottom); data observed in special days are highlighted in red.}
\label{fig:incre3}
\end{figure}

Before proceeding with the implementation of the predictive model, a suitable preprocessing step is performed in order to obtain a signal that can be forecast more effectively. A rather common step is resorting to a logarithmic transformation of the data \cite{nowicka2002modeling, incremona2020lassofft}, 
\begin{equation*}
S(d,q) := \ln\left(L(d, q)\right) 
\end{equation*}
which results in the time series displayed in Fig. \ref{fig:incre3} top. For a short-term prediction purpose, low frequency components such as trend and yearly periodicities can be neglected, while faster phenomena such as weekly seasonalities remain relevant and must be taken into account. In particular, weekly periodicity is modelled by assuming that, on a short range framework,
\begin{equation*}
S(d,q) = p(d, q) + \eta(d, q)
\end{equation*}
where $p(d, q)=p(d+7,q), \forall q$, is a deterministic periodic function in the first argument with period $T=7$ days and the time series $\eta(d, q)$ is a zero-mean stationary stochastic process, with the exception of the so called `intervention events', i.e. special days such as holidays, during which the typical weekly pattern is altered.
In order to filter the weekly periodicity of the signal, a 7-day differentiation is applied to the log-transformed time series:
\begin{equation*}
\tilde Y(d, q) := S(d,q) - S(d - 7,q)=\eta(d, q) - \eta(d-7, q)
\end{equation*} 

The resulting time series $\tilde{Y}(d, q)$, shown in Fig. \ref{fig:incre3} bottom, can be considered, in a short timespan, to be zero-mean and stationary, with the exception of special days that are excluded from this analysis, since they need an {\it{ad hoc}} modelling strategy.

In this preprocessing phase, the logarithmic transformation applied before the 7-day difference operator is crucial in order to make the marginal distributions of the data on each quarter-hour less skewed and closer to Gaussianity, which would allow an effective adoption of simple linear predictive models to achieve high forecasting performances.

Let $\mathpzc{D}_{s}$ be the set of all the special days (see Appendix) and $Y(d, q)$ the series of `cleaned' 7-day difference of log-load values, defined as:
\begin{equation*}
Y(d, q) = \begin{cases}  \mbox{missing}, & \mbox{if } (d \in \mathpzc{D}_{s})  \mbox{  or  }  (d-7 \in \mathpzc{D}_{s}) \\ \tilde{Y}(d, q), & \mbox{otherwise}  \end{cases}
\end{equation*}


\section{Problem statement} \label{sec:Problem_statement}

The main objective of this paper is the development of a one-day ahead forecaster $\hat{L}(d,q)$ for the daily profile of the Italian electric load $L(d, q), 1 \le q \le 96$, based on the knowledge of $L(t, q), \forall t < d, \forall q$. For the subsequent analysis, it is convenient to introduce the following lifted representation of the signals:
\begin{equation*}
\mathbf{L}(d) = \left[\begin{array}{cccc} L(d, 1) & L(d, 2) & \ldots & L(d, 96) \end{array} \right]^{T}
\end{equation*}
According to the lifted notation,
\begin{equation*}
\mathbf{L}(d) = \exp\left(\mathbf{\tilde{Y}}(d)\right) 
\end{equation*}
where exponentiation is applied elementwise.
\begin{equation*}
\mathbf{\tilde{Y}}(d) = \mathbf{S}(d) - \mathbf{S}(d - 7) 
\end{equation*}
and $\{\mathbf{Y} \}$ is a suitable subset of  $\{\mathbf{\tilde Y} \}$. The one-day ahead prediction problem then amounts to obtaining the prediction $\hat{\mathbf{L}}(d)$, based on $\{ \mathbf{L}(t), \; t < d \}$.

The solution approach will go through the calculation of a predictor $\mathbf{\hat{Y}}(d)$ of $\mathbf{Y}(d)$, given $\{ \mathbf{Y}(t), \; t < d \}$. Then, the predicted load is straightforwardly obtained as
\begin{equation*}
\mathbf{\hat{L}}(d) = \exp\left(\mathbf{\hat{S}}(d)\right) = \exp\left(\mathbf{\hat{Y}}(d) + \mathbf{S}(d-7)\right)
\end{equation*}
A general (nonlinear) prediction model can be written as 
\begin{equation*}
\mathbf{\hat Y}(d) = f\left(\mathbf{Y}(d-1), \mathbf{Y}(d-2), \ldots\right) 
\end{equation*}
where $f(\cdot,\cdot,\ldots)$ is a suitable nonlinear function. The predictor that minimizes the mean square error 
\begin{equation*}
\mathrm{MSE} = E\left[\left(\mathbf{\hat{Y}}(d) - \mathbf{Y}(d) \right)^{2} \bigg| \mathbf{Y}(d-1), \mathbf{Y}(d-2), \ldots \right]
\end{equation*}
is the conditional expectation
\begin{equation*}
f\left(\mathbf{Y}(d-1), \mathbf{Y}(d-2), \ldots\right) = E \left[\mathbf{Y}(d) \vert \mathbf{Y}(d-1), \mathbf{Y}(d-2), \ldots  \right]
\end{equation*}
Estimating the conditional expectation is generally a demanding task, but a dramatic simplification occurs when the stochastic process $\left\{ \mathbf{Y}(\cdot)\right\}$ is Gaussian, in which case the conditional expectation is a linear function of past observations:
\begin{equation*}\label{eq:conditional_expectation}
E \left[\mathbf{Y}(d) \vert \mathbf{Y}(d-1), \ldots , \mathbf{Y}(d-n) \right]
= \sum_{i=1}^{n}\mathbf{A}_{i}\mathbf{Y}(d-i)
\end{equation*}

where $\mathbf{A}_{i}$ are suitable coefficient matrices which depend on the second order statistics of the process $\mathbf{Y}(\cdot)$ and will have to be estimated from data.

\begin{figure}
\centering
\includegraphics[width=.8\textwidth]{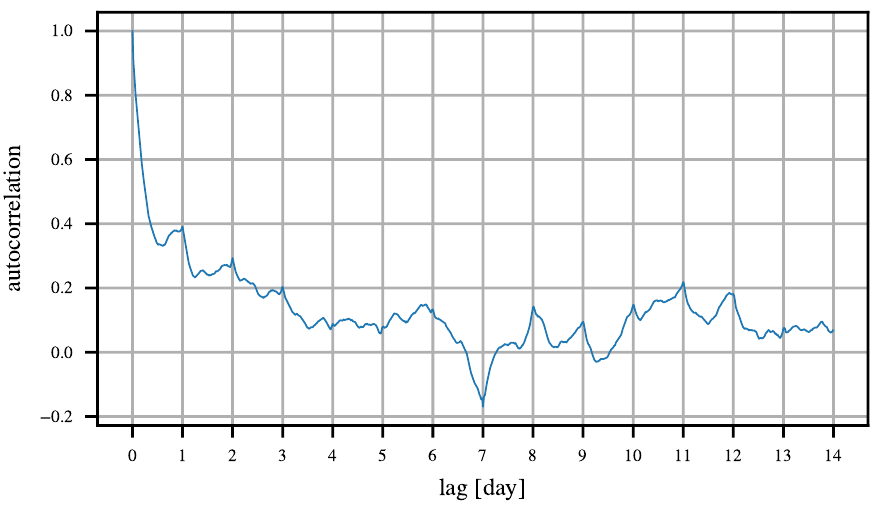}
\caption{Autocorrelation of 7-day difference of log-load values computed on the first week of October 2019. As a consequence of the 7-day difference, the autocorrelation tends to stay close to zero as the lag approaches 7 days.}
\label{fig:incre4}
\end{figure}

The next step is the choice of the order $n$ of the predictor. In Fig. \ref{fig:incre4} it is possible to see that the autocorrelation of the signal, as a consequence of the 7-day differentiation, converges to zero rather quickly, suggesting that a low-order predictor may suffice.

In particular the most simple model will be the first-order one
\begin{equation} \label{eq:model_equation_1}
\mathbf{\hat Y}(d) = \mathbf{A}\mathbf{Y}(d-1)
\end{equation}
where $\mathbf{A}$ is a $96\times96$ matrix of weights defined as:
\begin{equation*}
\mathbf{A} =
\begin{bmatrix}
a_{1,1} & a_{1,2} & \dots & a_{1,96} \\
a_{2,1} & a_{2,2} & \ddots & \vdots \\
\vdots & \ddots & \ddots & \vdots \\
a_{96,1} & \dots & \ldots & a_{96,96}  \\
\end{bmatrix}
\end{equation*}

This predictor structure lends itself to an insightful interpretation. In fact, $a_{i,j}$ is the weight assigned to $\mathbf{Y}(d-1)_j=Y(d-1, j)$ for predicting $\mathbf{Y}(d)_i=Y(d,i)$. With these assumptions, the purpose of this work will be the estimation of the weights $\{a_{i,j}\}$.


\section{Forecasting methods} \label{sec:Forecasting_methods}

Let
\begin{equation*}
\mathbf{a} = \vect(\mathbf{A}) = \left[\begin{array}{ccccccc} a_{1,1} & \ldots & a_{1,96} & \ldots & a_{96,1} & \ldots & a_{96,96}\end{array}\right]^{T} \in R^{96^2}
\end{equation*}
be the vectorization of $\mathbf{A} \in R^{96 \times 96}$. Moreover, let
\begin{equation*}
  \mathbf{y} =
  \left[ {\begin{array}{c}
   \mathbf{Y}(1)  \\
   \mathbf{Y}(2) \\
    \vdots \\
    \mathbf{Y}(n_{day})
  \end{array} } \right]
\in R^{96 n_{day}} 
\end{equation*}
be the vector of all outputs and

\begin{equation*}
  \mathbf{\Phi} =
  \left[ {\begin{array}{c}
   \mathbf{I}_{96\times96} \otimes \mathbf{Y}^{T}(0)  \\
    \mathbf{I}_{96\times96} \otimes \mathbf{Y}^{T}(1) \\
    \vdots \\
    \mathbf{I}_{96\times96} \otimes \mathbf{Y}^{T}(n_{day} - 1)  \\
  \end{array} } \right]
\in R^{96 n_{day}\times 96^2}
\end{equation*}
the regressor matrix, where $n_{day}$ is the number of considered days.

Then, letting $\mathbf{\hat{y}}$ denote the prediction of $\mathbf{{y}}$, it is possible to rewrite \eqref{eq:model_equation_1} as follows:
\begin{equation*}
\mathbf{\hat{y}} = \mathbf{\Phi} \mathbf{a} \label{eq:model_equation_2}
\end{equation*}

In the following, six different techniques are considered for the estimation of $\mathbf{a}$: Ordinary Least Squares (OLS), a Tikhonov-based amplitude regularization model (TA), a Tikhonov-based second derivative regularization model (TS), a Regularized Radial Basis Functions-based model (RBF), and two sparse models selecting just a suitable subset of the weights, the `Two-Edges' (TE) and the `One-Edge' (OnE) models.

\subsection{Ordinary Least Squares approach (OLS)}
The most direct way to estimate the weight vector $\mathbf{a}$ is to resort to ordinary least squares:
\begin{equation}
\mathbf{a}^{OLS} = \argmin_{\mathbf{a}}(\mathbf{y} - \mathbf{\Phi}\mathbf{a})^{T}(\mathbf{y} - \mathbf{\Phi}\mathbf{a}) \label{eq:least_squares_opt}
\end{equation}

Recall that each quarter-hour of the target day is predicted as the linear combination of all the quarter-hours of the previous day. Then, it is easy to see that \eqref{eq:least_squares_opt} is completely equivalent to  $96$ OLS problems, each of which provides the OLS estimate of one of the $96$ rows of $\mathbf{A}$. This approach is consistent with the multimodel paradigm to the joint design of predictors for different horizons \cite{hong1998weighted}, \cite{ahmia2015multimodel}. The multimodel paradigm offers a flexible alternative to the single-model approach that relies on a unique model of a stochastic process from which multistep optimal predictors are computed. However, when a unique model is estimated from data it may suffer from some bias that propagates to the predictors. For instance, if a Prediction Error Method is used for identifying the model, the one-step-ahead predictor errors are minimized, but, if the model is biased, there is no guarantee that long-range predictions are equally satisfactory. Hence the idea of estimating a different predictor for each prediction range, which goes under the name of multi-model approach. In this way, it is possible to reduce the bias of each single predictor, because more degrees of freedom are available. This is obviously more flexible at the cost of possible overparametrization. In our case, in fact, we are estimating $96 \times 96 = 9216$ independent parameters.

\begin{figure*}
  \centering 
  \subfigure[3D view]{\includegraphics[scale=0.225]{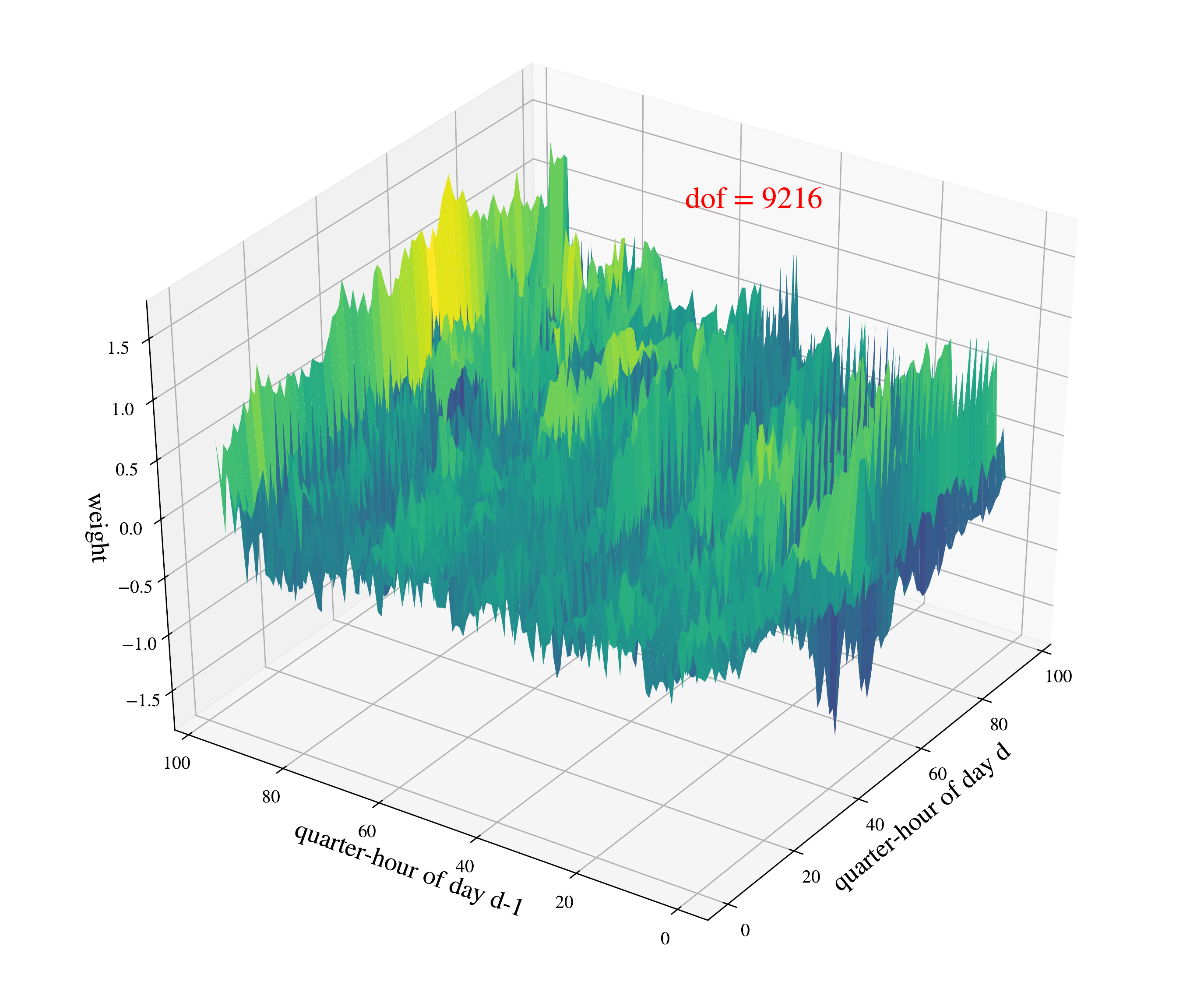}}
  \subfigure[Top view]{\includegraphics[scale=0.285]{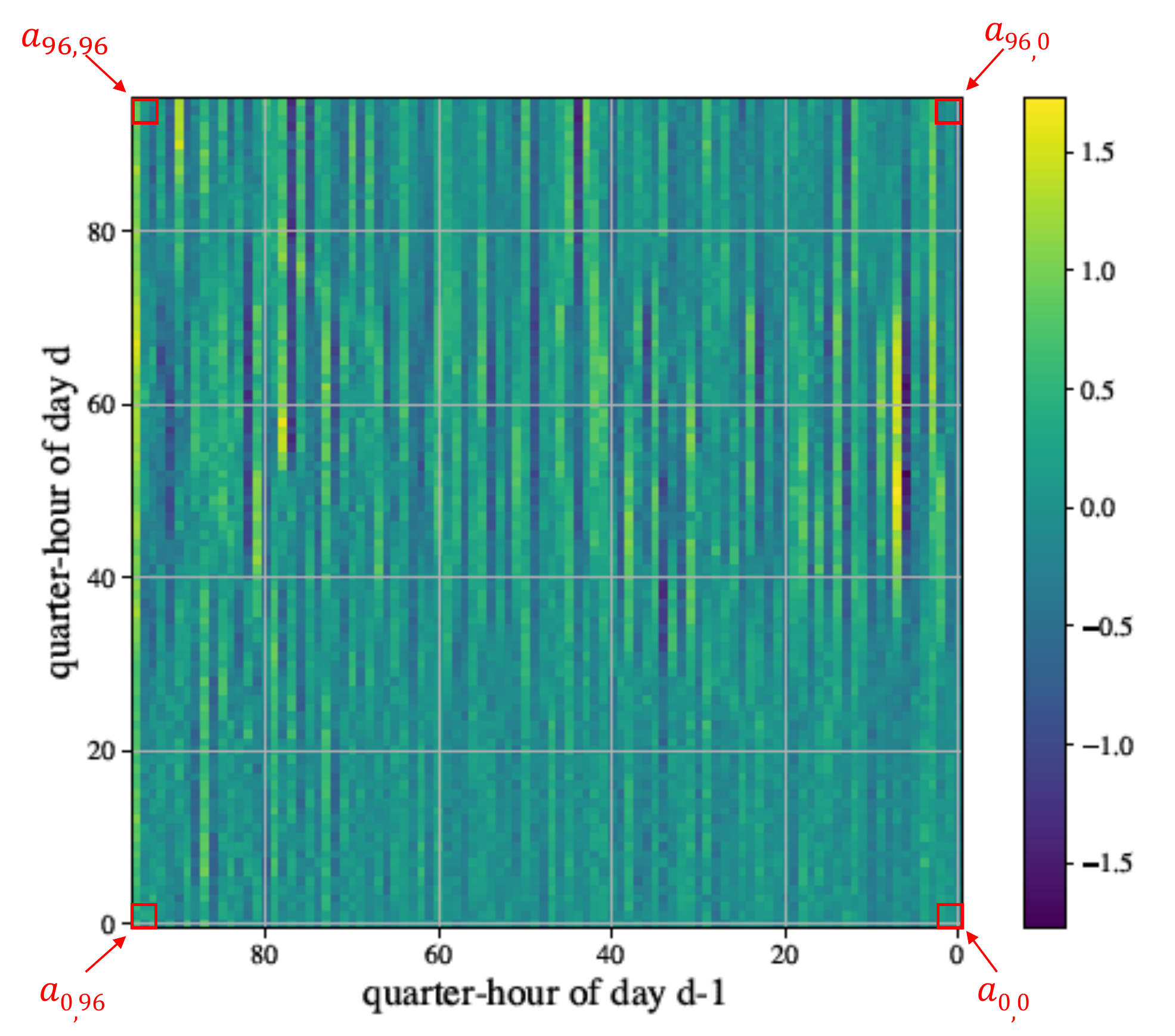}} 
\caption{Ordinary Least Squares approach weight surface estimated on the 2018 data: 3D view (left) and top view (right).}
\label{fig:incre5}
\end{figure*}

In view of the previous considerations, it is not surprising that, when we display as a surface the entries of matrix $\mathbf{A}$ estimated from 2018 data, it turns out to be very rough, see Fig. \ref{fig:incre5}. The roughness reflects two features: the variance of the estimates and the oscillations from one column to another. This last feature is best appreciated by looking at the top view displayed in Panel (b) of Fig. \ref{fig:incre5}, where the colormap exhibits vertical stripes, a symptom of greater variability across columns than across rows. 

This different variability can be explained by considering the problem of predicting the target (i.e. the seven-day difference of the log-loads) at two consecutive quarter-hours $i$ and $i+1$, i.e. the problem of predicting $Y(d,i)$ and $Y(d,i+1)$, given $\mathbf{Y}(d-1)$. The log-loads are sampled frequently and cannot vary abruptly from one quarter-hour to another, a property that propagates to the seven-day difference, so that $Y(d,i) \approx Y(d,i+1)$. Observe also that the two predictors 
\begin{eqnarray*}
\hat Y(d,i) &=& \sum_{j=1}^{96}\mathbf{a}_{i,j}Y(d-1,j) \\
\hat Y(d,i+1) &=& \sum_{j=1}^{96}\mathbf{a}_{i+1,j}Y(d-1,j)
\end{eqnarray*}
share the same regressors, i.e. the vector $\mathbf{Y}(d-1)$.
Since
\begin{equation*}
\hat Y(d,i) \approx \hat Y(d,i+1)
\end{equation*}
it follows that, for each given $j$, the weights $\mathbf{a}_{i,j}$ and $\mathbf{a}_{i+1,j}$ cannot be too different, which explains the smaller variability across rows.

The irregularity of the  surface derives from the overparametrization of the model. On the other hand, there are good reasons for the weight surface to be smooth. Indeed, for any given $i$, it is reasonable to assume that weights $\mathbf{a}_{i,j}$ and $\mathbf{a}_{i,j+1}$, associated to consecutive quarter-hours, do not differ very much from each other. This justifies the design of alternative estimation schemes that reduce the degrees of freedom by enforcing some kind of smoothness on the weight surface.

\subsection{Tikhonov regularization}
The shortcomings of the OLS estimate can be addressed through Tikhonov regularization techniques which, at the cost of some bias, reduce the variance (and the degrees of freedom) of the estimate, by adding a penalty term to the quadratic loss function \eqref{eq:least_squares_opt} \cite{gruber2017shrinkage, saleh2019ridge}. For the problem of predicting $\mathbf{y}$ by means of $\mathbf{\hat y}= \mathbf{X} \bm{\beta}$, the Tikhonov estimate of $\bm{\beta}$ is 
\begin{equation}
\bm{\beta}^{reg} = \argmin_{\bm{\beta}}(\mathbf{y} - \mathbf{X}\bm{\beta})^{T}(\mathbf{y} - \mathbf{X}\bm{\beta}) +  \bm{\beta}^{T} \mathbf{T} \bm{\beta}  \label{eq:tikhonov_reg_opt}
\end{equation}
where $\mathbf{T}>0$ is a matrix whose choice determines the type of regularization. For instance, $\mathbf{T}=\lambda\mathbf{\Gamma}^{T}\mathbf{\Gamma}$ yields
\begin{equation*}
\bm{\beta}^{T} \mathbf{T} \bm{\beta} = \lambda \Vert \mathbf{\Gamma} \bm{\beta} \Vert_{2}^{2}
\end{equation*}
where $\mathbf{\Gamma}$ is the \emph{Tikhonov matrix} and $\lambda$ is a regularization parameter that controls the balance between the residual sum of squares and the penalty term in \eqref{eq:tikhonov_reg_opt}. The tuning of $\lambda$ can be performed according to different methods. Hereafter, a cross-validation approach is adopted, whose details are given in Section \ref{sec:Experimental_validation_setup}.

The solution to \eqref{eq:tikhonov_reg_opt} is 
\begin{equation*}
\bm{\beta}^{reg} = \left(\mathbf{X}^{T}\mathbf{X}+\mathbf{T} \right)^{-1} \mathbf{X}^{T}\mathbf{y}
\label{eq:tikhonov_solution}
\end{equation*}
so that
\begin{equation*}
\mathbf{\hat y}= \mathbf{H}\mathbf{y},\quad
\mathbf{H} =\mathbf{X} \left(\mathbf{X}^{T}\mathbf{X}+\mathbf{T} \right)^{-1} \mathbf{X}^{T}
\end{equation*}
where $\mathbf{H}$ is the so-called `hat matrix'. A useful measure of the complexity of the model is given by the equivalent degrees of freedom, defined as
\begin{equation*}
\mathrm{dof}= \Tr \left(\mathbf{H}\right).
\end{equation*}

\paragraph{Tikhonov amplitude regularization (TA)}
A possible regularization strategy consists of applying a penalty to the amplitude of the parameters in order to favor solutions with smaller norm. This technique, also known as \emph{ridge regression}, is associated with the following choice of the $\mathbf{T}$ matrix:
\begin{equation*}
\mathbf{T} = \lambda \mathbf{I}.
\end{equation*}

\begin{figure*}
  \centering 
  \subfigure[3D view]{\includegraphics[scale=0.223]{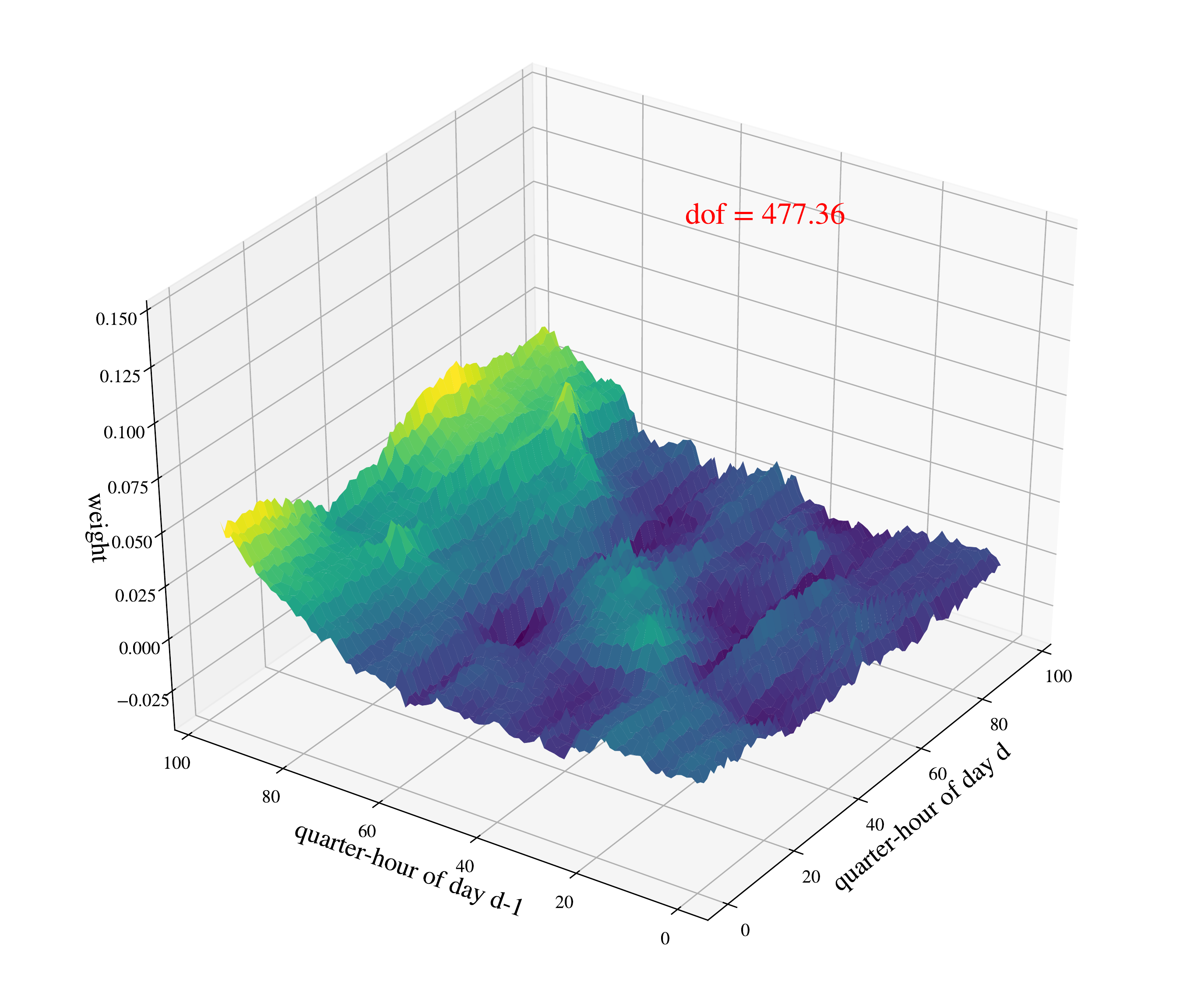}}
  \subfigure[Top view]{\includegraphics[scale=0.283]{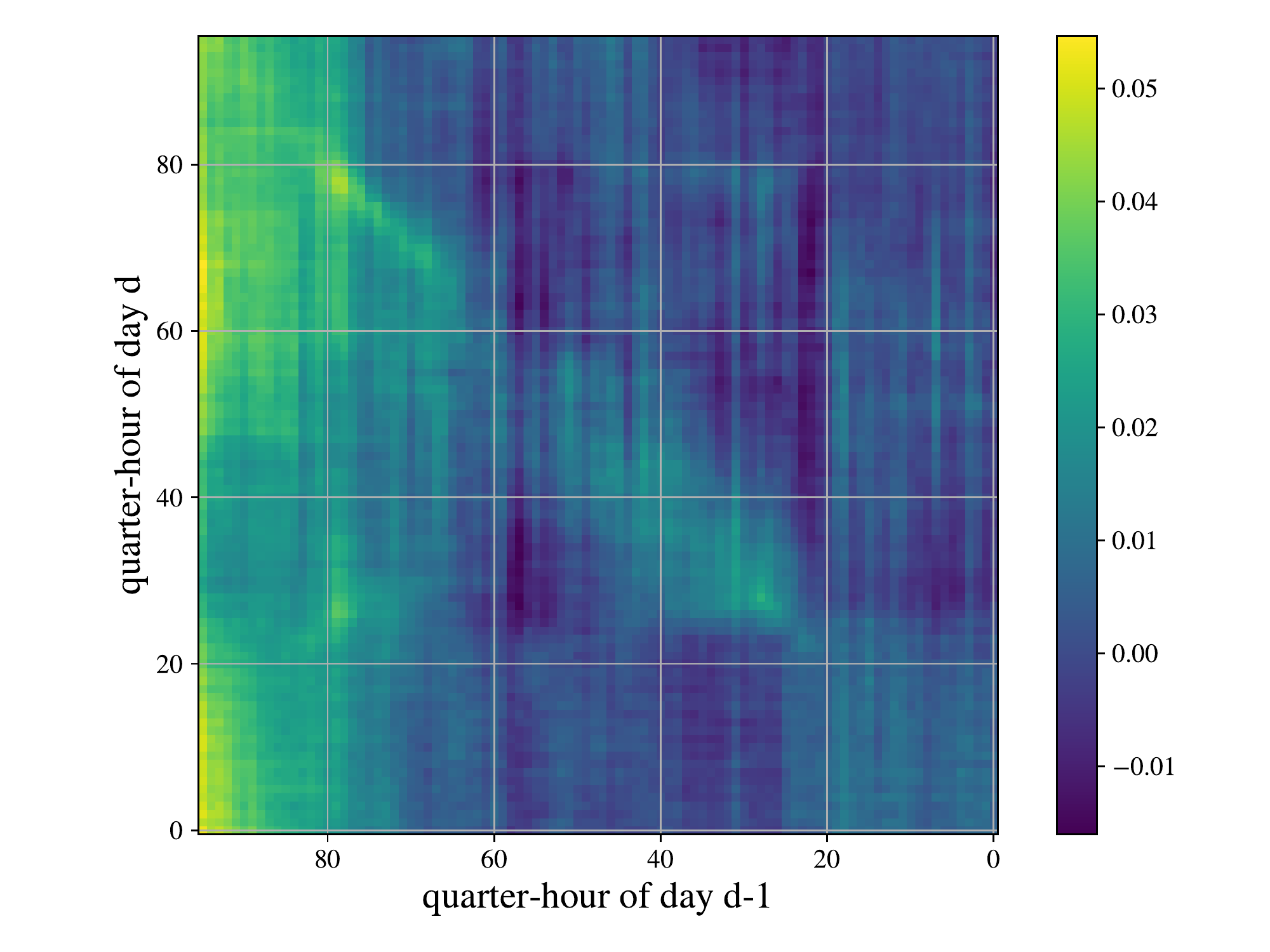}} 
\caption{Tikhonov amplitude (TA) regularized weight surface estimated on the 2018 data: 3D view (left) and top view (right). In the latter view the yellow and light green colors highlight the presence of two main edges, a vertical and a diagonal one.}
\label{fig:incre6}
\end{figure*}

Letting $\mathbf{X}=\mathbf{\Phi}$ and $\bm{\beta}=\mathbf{a}$, the corresponding regularized weight surface estimated from the 2018 data is displayed in Fig. \ref{fig:incre6}. Compared to Fig. \ref{fig:incre5}, the shape is smoother, although some `stripe effect' is still visible in Panel (b). Indeed, it is easy to see that solving \eqref{eq:tikhonov_reg_opt} with $\mathbf{T} = \lambda \mathbf{I}$ is equivalent to solving $96$ independent ridge regression problems, one for each row of $\mathbf{A}$. In other words, regularity is enforced by damping the amplitude of the entries of $\mathbf{A}$, but oscillations between columns are not explicitly penalized.

It is worth noting that the visual inspection of the Top view (Panel (b) of Fig. \ref{fig:incre6}) reveals the presence of two `edges', one horizontal and one diagonal, highlighted by the yellow/light green color. The former is in correspondence with the left side of the square (associated with the last column of matrix $\mathbf{A}$) and the latter is in correspondence of the diagonal of the square (associated with the main diagonal of matrix $\mathbf{A}$). This observation will be the basis of two regularization methods (OnE and TE) that reduce the degrees of freedom by imposing a structured sparse structure on $\mathbf{A}$.

The two edges, that identify the most relevant regressors for the prediction of tomorrow's target variable, admit a very meaningful interpretation. The vertical edge highlights the importance of the most recent observations, i.e. those just before midnight. The diagonal edge, conversely, indicates that, when predicting tomorrow's $i$-th value $Y(d,i)$, a great weight is assigned to $Y(d-1,i)$, which is rather intuitive.

\paragraph{Tikhonov second derivative regularization (TS)} The idea of this approach is to force the weight surface to be `smooth' along both directions. In order to do so, two penalty terms are applied to \eqref{eq:least_squares_opt} in order to penalize the squares of the second differences along both rows and columns. Letting $\mathbf{X}=\mathbf{\Phi}$ and $\bm{\beta}=\mathbf{a}$, the corresponding Tikhonov regularization problem can be stated as follows:
\begin{equation}
\mathbf{a}^{der2} = \argmin_{\mathbf{a}}(\mathbf{y} - \mathbf{\Phi}\mathbf{a})^{T}(\mathbf{y} - \mathbf{\Phi}\mathbf{a}) + \lambda_{1} \Vert 
\mathbf{\Delta}_{1} \mathbf{a} \Vert_{2}^{2} + \lambda_{2} \Vert \mathbf{\Delta}_{2} \mathbf{a} \Vert_{2}^{2}
\label{eq:tikhonov_second_derivative}
\end{equation}
where $\lambda_{1}$ and $\lambda_{2}$ are two regularization parameters and $\mathbf{\Delta}_{1} \in R^{94 \cdot 96\times 96^2}$ and $\mathbf{\Delta}_{2}\in R^{94 \cdot 96\times 96^2}$ are such that
\begin{equation*}
\mathbf{\Delta}_{1}\mathbf{a} = 
\begin{bmatrix}
a_{1,3} + 2a_{1,2} - a_{1,1} \\
a_{1,4} + 2a_{1,3} - a_{1,2} \\
\vdots \\
a_{1,96} + 2a_{1,95} - a_{1,94} \\
a_{2,3} + 2a_{2,2} - a_{2,1} \\
\vdots \\
a_{96,96} + 2a_{96,95} - a_{96,94}
\end{bmatrix},
\quad
\mathbf{\Delta}_{2}\mathbf{a} = 
\begin{bmatrix}
a_{3,1} + 2a_{2,1} - a_{1,1} \\
a_{4,1} + 2a_{3,1} - a_{2,1} \\
\vdots \\
a_{96,1} + 2a_{95,1} - a_{94,1} \\
a_{3,2} + 2a_{2,2} - a_{1,2} \\
\vdots \\
a_{96,96} + 2a_{95,96} - a_{94,96}
\end{bmatrix}
\end{equation*}
yield the row-wise and column-wise second differences of the entries of $\mathbf{A}$. It is immediate to see that the corresponding $\mathbf{T}$ matrix is
\begin{equation*}
\mathbf{T} = \lambda_{1} \mathbf{\Delta}^{T}_{1}\mathbf{\Delta}_{1} + \lambda_{2}\mathbf{\Delta}^{T}_{2}\mathbf{\Delta}_{2}.
\end{equation*}
It is worth noting that if $\lambda_{1}=\lambda_{2}$ the regularization penalty in \eqref{eq:tikhonov_second_derivative} boils down to the classical discrete Laplacian operator. In this work, the formulation with two independent regularization parameters is preferred in view of its greater flexibility.

\begin{figure*}
  \centering 
  \subfigure[3D view]{\includegraphics[scale=0.223]{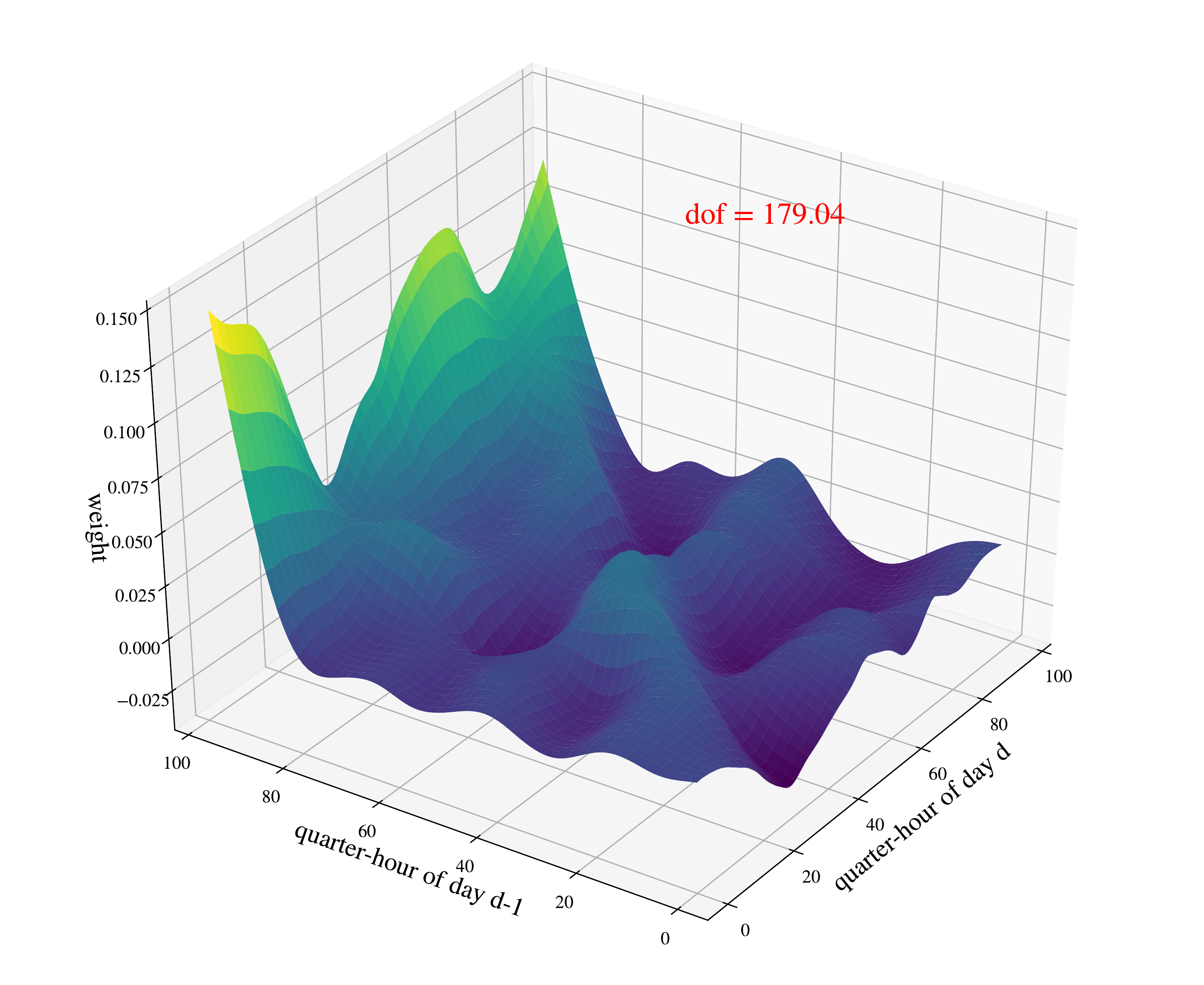}}
  \subfigure[Top view]{\includegraphics[scale=0.283]{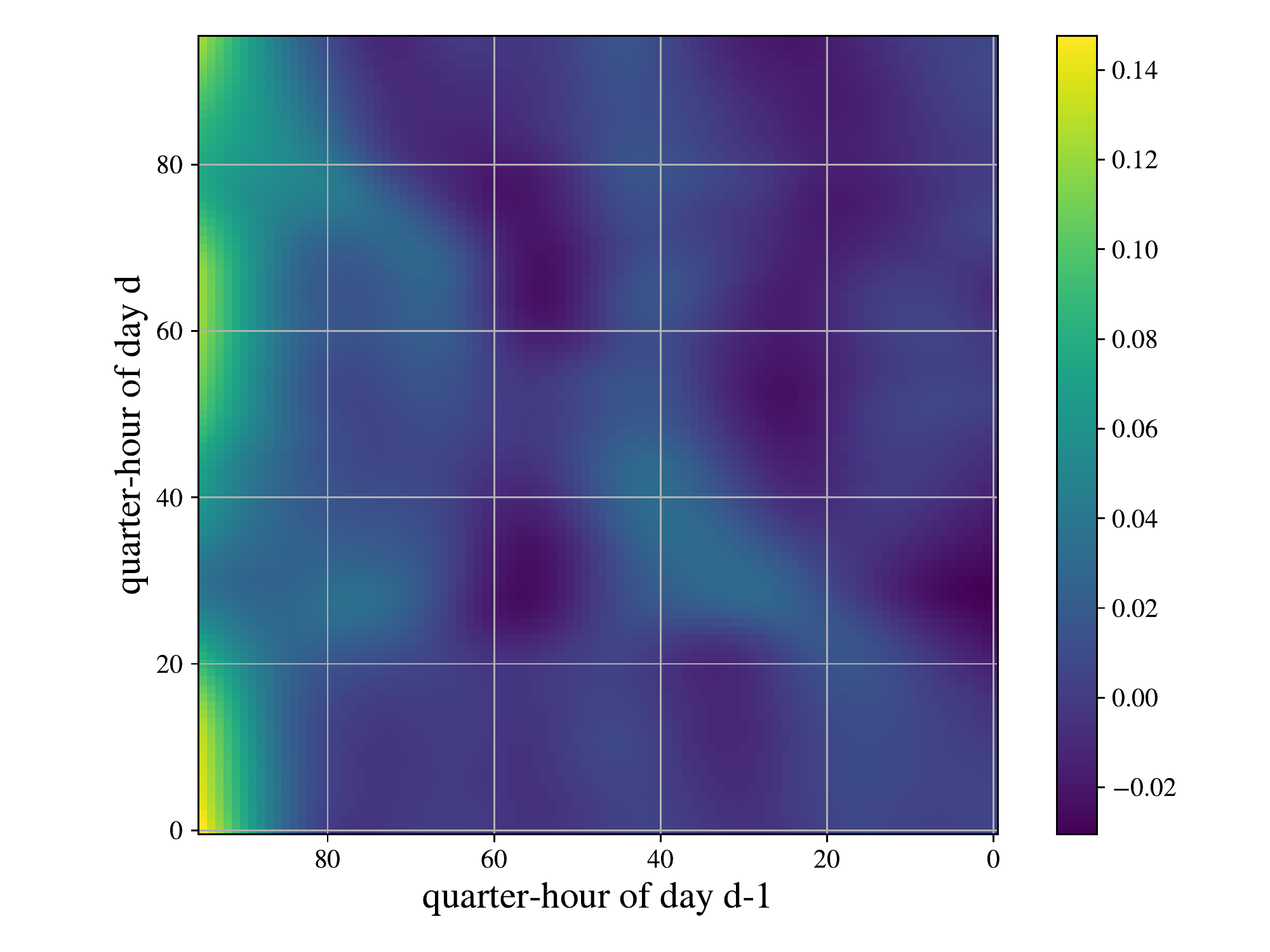}} 
\caption{Tikhonov second derivative regularized weight surface estimated on the 2018 data: 3D view (left) and top view (right).}
\label{fig:incre7}
\end{figure*}

The corresponding surface, displayed in Fig. \ref{fig:incre7}, is even smoother than the TA one. The vertical and diagonal edges are still well seen in Panel (b). 

\subsection{Regularized Radial Basis Functions regularization (RBF)}
\label{sec:rbf_subsection}
The approach described in this subsection consists of regularizing the weight surface by representing it as the sum of a cubic polynomial term, which is used to capture its trends, and a set of Regularized Radial Basis Functions (RBF) to capture the local details of the surface \cite{ranaweera1995rbf}, \cite{konishi2004rbf}. In particular:
\begin{equation*}
a_{i,j} = \bar{a}_{i,j} + \tilde{a}_{i,j}
\end{equation*}
\begin{equation*}
\bar{a}_{i,j} = c_{1} + c_{2}i + c_{3}j + c_{4}i^{2} + c_{5}ij + c_{6}j^{2} + c_{7}i^{3} + c_{8}i^{2}j + c_{9}ij^{2} + c_{10}j^{3}
\end{equation*}
\begin{equation*}
\tilde{a}_{i,j} = \sum_{k=0}^{m}\sum_{z=0}^{m}\theta_{k,z}\phi\left( \sqrt{\left(i - w_{k}\right)^{2} + \left(j - v_{z}\right)^{2}} \right)
\end{equation*}
where 
\begin{equation*}
\phi\left( r \right) = \exp\left((-r^{2})/(2\sigma^{2})\right)
\end{equation*}

and $\left(w_{k}, v_{z}\right)$, with $k,z=1,...,m$, are the coordinates of the centers of the radial functions that are assumed to be located on a uniform square grid:
\begin{eqnarray}
w_{k}&=&\frac{96 k}{m}, \quad k=0, \ldots, m \\
w_{z}&=&\frac{96 z}{m}, \quad z=0, \ldots, m
\end{eqnarray}

and $\sigma$ is the standard deviation.

Notice that the parameter vector $\mathbf{a}$ can be written as
\begin{equation*}
\mathbf{a} = \mathbf{\bar{a}} + \mathbf{\tilde{a}}
\end{equation*}

with the corresponding weight surface given by
\begin{equation*}
\mathbf{A} = \mathbf{\bar{A}} + \mathbf{\tilde{A}}
\end{equation*}
where $\mathbf{\bar{A}}$ denotes the polynomial component and $\mathbf{\tilde{A}}$ the RBF one.
Once again, the optimization problem is the one given in \eqref{eq:tikhonov_reg_opt}, with
\begin{equation*}
\mathbf{y} = \mathbf{\Phi} ( \mathbf{\bar a} +\mathbf{\tilde a}),\quad \mathbf{\bar a} = \mathbf{P} \mathbf{c}, \quad \mathbf{\tilde a}= \mathbf{R} \bm{\theta}
\end{equation*}
so that
\begin{equation*}
\mathbf{X}=\mathbf{\Phi} \left[\begin{array}{cc} \mathbf{P} & \mathbf{R} \end{array} \right], \quad \bm{\beta} = \left[\begin{array}{cc} \mathbf{c} \\ \bm{\theta} \end{array} \right]
\end{equation*}

where the matrices $\mathbf{P}$ and $\mathbf{R}$ are the cubic polynomial and the radial basis functions matrices, respectively. The matrix $\mathbf{T}$ applies ridge regularization to the amplitudes of the radial basis functions, while no shrinking is applied to the polynomial coefficients.

\begin{equation*}
\mathbf{T} =
\begin{bmatrix}
\mathbf{0} & \mathbf{0} \\
\mathbf{0} & \lambda\mathbf{I_m} 
\end{bmatrix}
\end{equation*}
In particular,
\begin{equation*}
  \mathbf{P}=
  \left[ {\begin{array}{cccccc}
   \mathbf{1} & \mathbf{i} & \mathbf{j} & \mathbf{i^2} & \ldots & \mathbf{j^3}  \\
  \end{array} } \right]
  \in R^{96^2 \times 10}
\end{equation*}

\begin{equation*}
  \mathbf{R}=
  \left[ {\begin{array}{ccc}
   \phi\left( \bigg\Vert \begin{bmatrix} 1 - w_{0} \\
   1 - v_{0} \end{bmatrix}\bigg\Vert_{2} \right) & \ldots & \phi\left( \bigg\Vert \begin{bmatrix} 1 - w_{m} \\
   1 - v_{m} \end{bmatrix}\bigg\Vert_{2} \right) \\
    \vdots &\vdots & \vdots \\
    \phi\left( \bigg\Vert \begin{bmatrix} 1 - w_{0} \\
   96 - v_{0} \end{bmatrix}\bigg\Vert_{2} \right) & \ldots & \phi\left( \bigg\Vert \begin{bmatrix} 1 - w_{m} \\
   96 - v_{m} \end{bmatrix}\bigg\Vert_{2} \right) \\ \\

    \phi\left( \bigg\Vert \begin{bmatrix} 2 - w_{0} \\
   1 - v_{0} \end{bmatrix}\bigg\Vert_{2} \right) & \ldots & \phi\left( \bigg\Vert \begin{bmatrix} 2 - w_{m} \\
   1 - v_{m} \end{bmatrix}\bigg\Vert_{2} \right) \\
    \vdots &\vdots & \vdots \\
    \phi\left( \bigg\Vert \begin{bmatrix} 2 - w_{0} \\
   96 - v_{0} \end{bmatrix}\bigg\Vert_{2} \right) & \ldots & \phi\left( \bigg\Vert \begin{bmatrix} 2 - w_{m} \\
   96 - v_{m} \end{bmatrix}\bigg\Vert_{2} \right) \\

    \vdots & \vdots & \vdots \\
    \phi\left( \bigg\Vert \begin{bmatrix} 96 - w_{0} \\
   1 - v_{0} \end{bmatrix}\bigg\Vert_{2} \right) & \ldots & \phi\left( \bigg\Vert \begin{bmatrix} 96 - w_{m} \\
   1 - v_{m} \end{bmatrix}\bigg\Vert_{2} \right) \\
    \vdots & \vdots & \vdots \\
    \phi\left( \bigg\Vert \begin{bmatrix} 96 - w_{0} \\
   96 - v_{0} \end{bmatrix}\bigg\Vert_{2} \right)  & \ldots & \phi\left( \bigg\Vert \begin{bmatrix} 96 - w_{m} \\
   96 - v_{m} \end{bmatrix}\bigg\Vert_{2} \right) \\
  \end{array} } \right]
  \in R^{96^2 \times m^2}
\end{equation*}

where
\begin{equation*}
\mathbf{i} = \left[\begin{array}{cccccccccc} 1 & \ldots & 96 & 1 & \ldots & 96 & \ldots & 1 & \ldots & 96 \end{array}\right]^{T} \in R^{96^2 \times 1},
\end{equation*}
\begin{equation*}
\mathbf{j} = \left[\begin{array}{cccccccccc} 1 & \ldots & 1 & 2 & \ldots & 2 & \ldots & 96 & \ldots & 96 \end{array}\right]^{T} \in R^{96^2 \times 1}
\end{equation*}

\begin{equation*}
  \mathbf{c}=
  \left[ {\begin{array}{cccc}
   c_{1} & c_{2} & \dots & c_{10} \\
  \end{array} } \right]^{T}
\in R^{10 \times 1}
\end{equation*}
\begin{equation*}
  \bm{\theta}=
  \left[ {\begin{array}{cccccc}
   \theta_{1,1} &
   \dots &
   \theta_{1,m} &
   \theta_{2,1} &
   \dots &
   \theta_{m,m} \\
  \end{array} } \right]^{T}
\in R^{m^2 \times 1}
\end{equation*}

In this work, the value of $\sigma$ has been fixed to $4$ while $m$ has been fixed to $12$ (which leads to $13\times13=169$ bell-shaped basis functions). The vector $\theta$ contains the parameters $\theta_{k,z}$, each one representing the amplitude of the radial function centered in $\left(w_{k},v_{z}\right)$.

\begin{figure*} 
  \centering 
  \subfigure[Cubic polynomial component]{\includegraphics[scale=0.223]{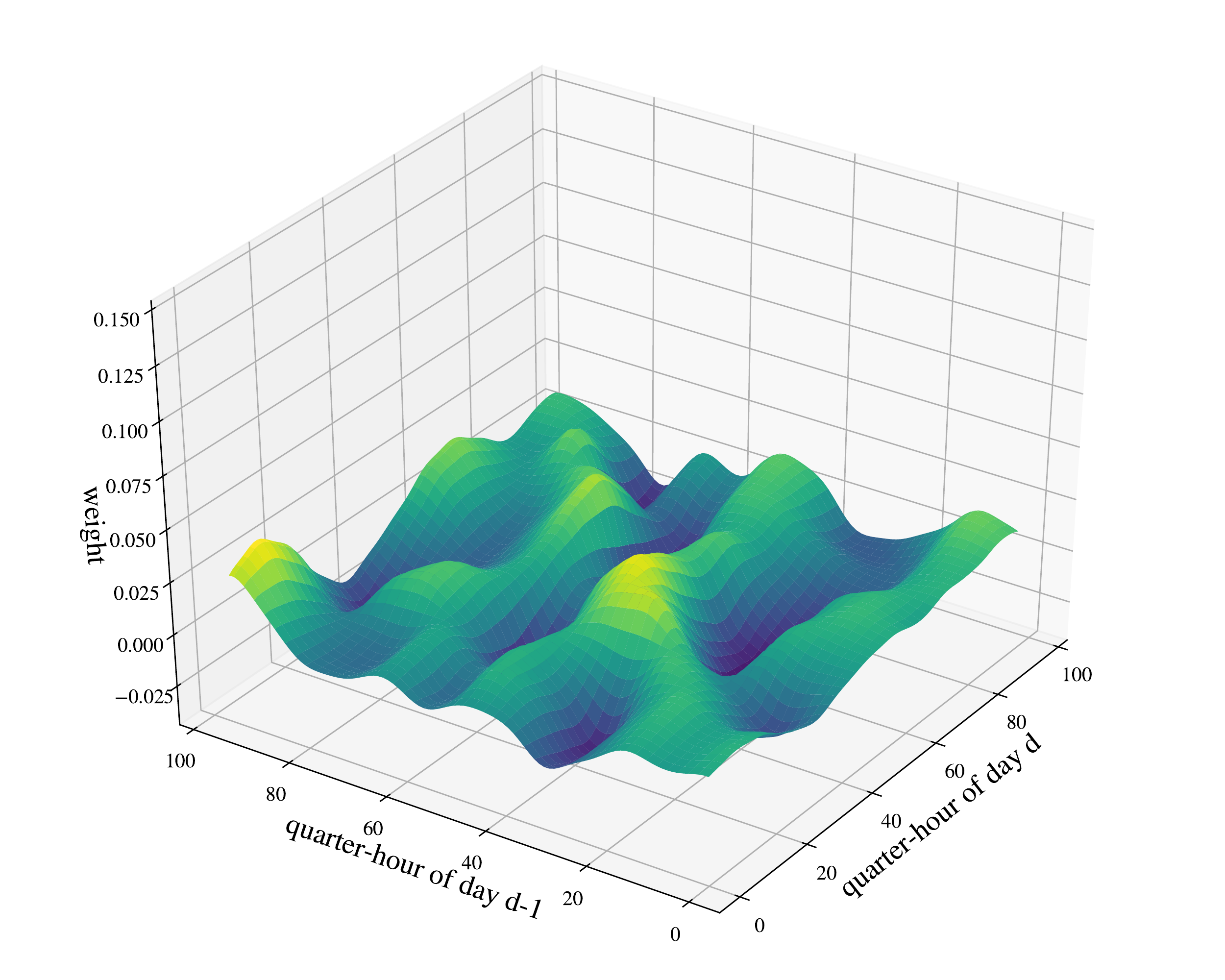} \label{fig:incre8topleft}}
  \subfigure[Gaussian component]{\includegraphics[scale=0.223]{incre8topright.pdf} \label{fig:incre8topright}}
  \subfigure[Regularized RBF surface (3D view)]{\includegraphics[scale=0.223]{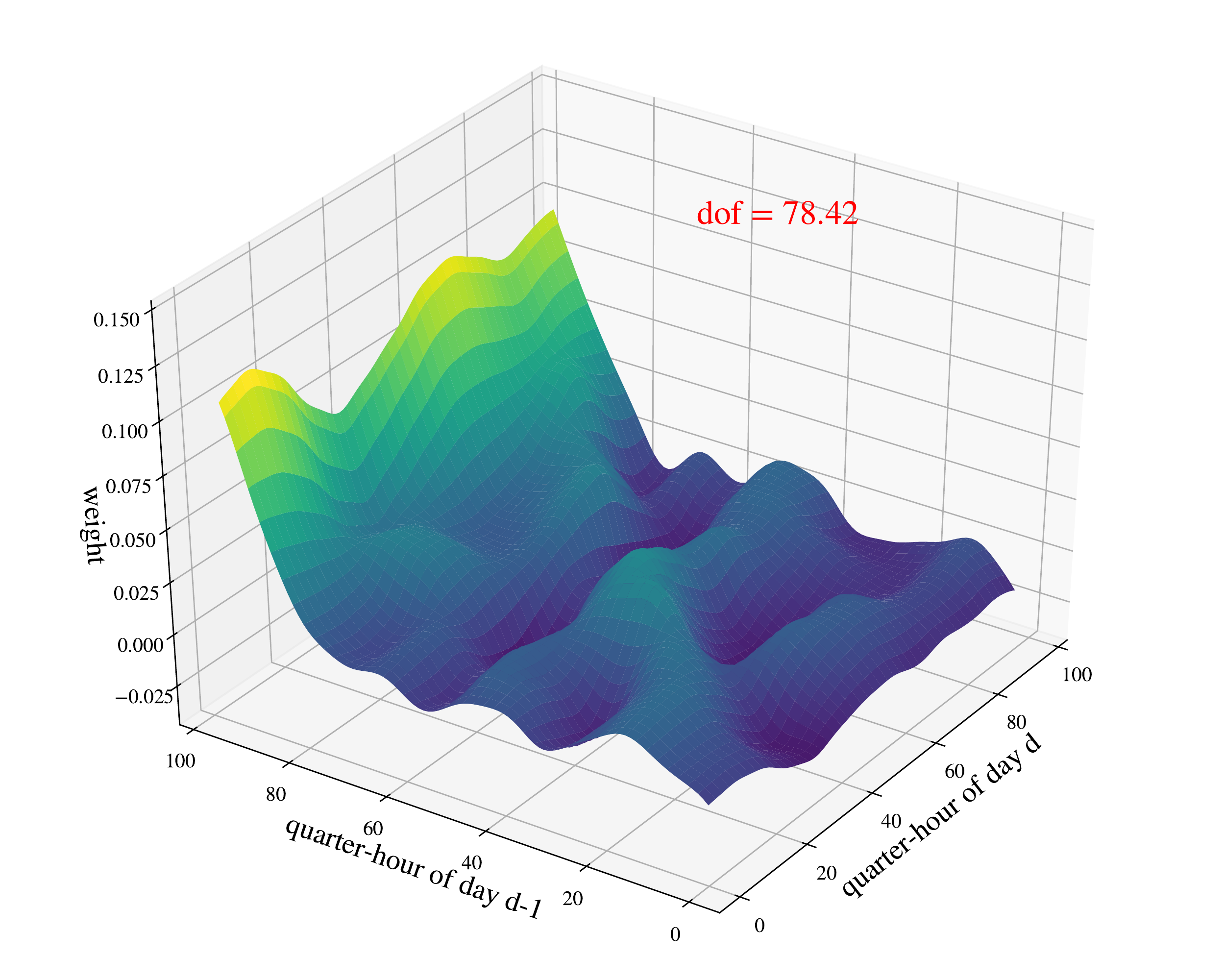} \label{fig:incre8bottomleft}}
  \subfigure[Regularized RBF surface (Top view)]{\includegraphics[scale=0.283]{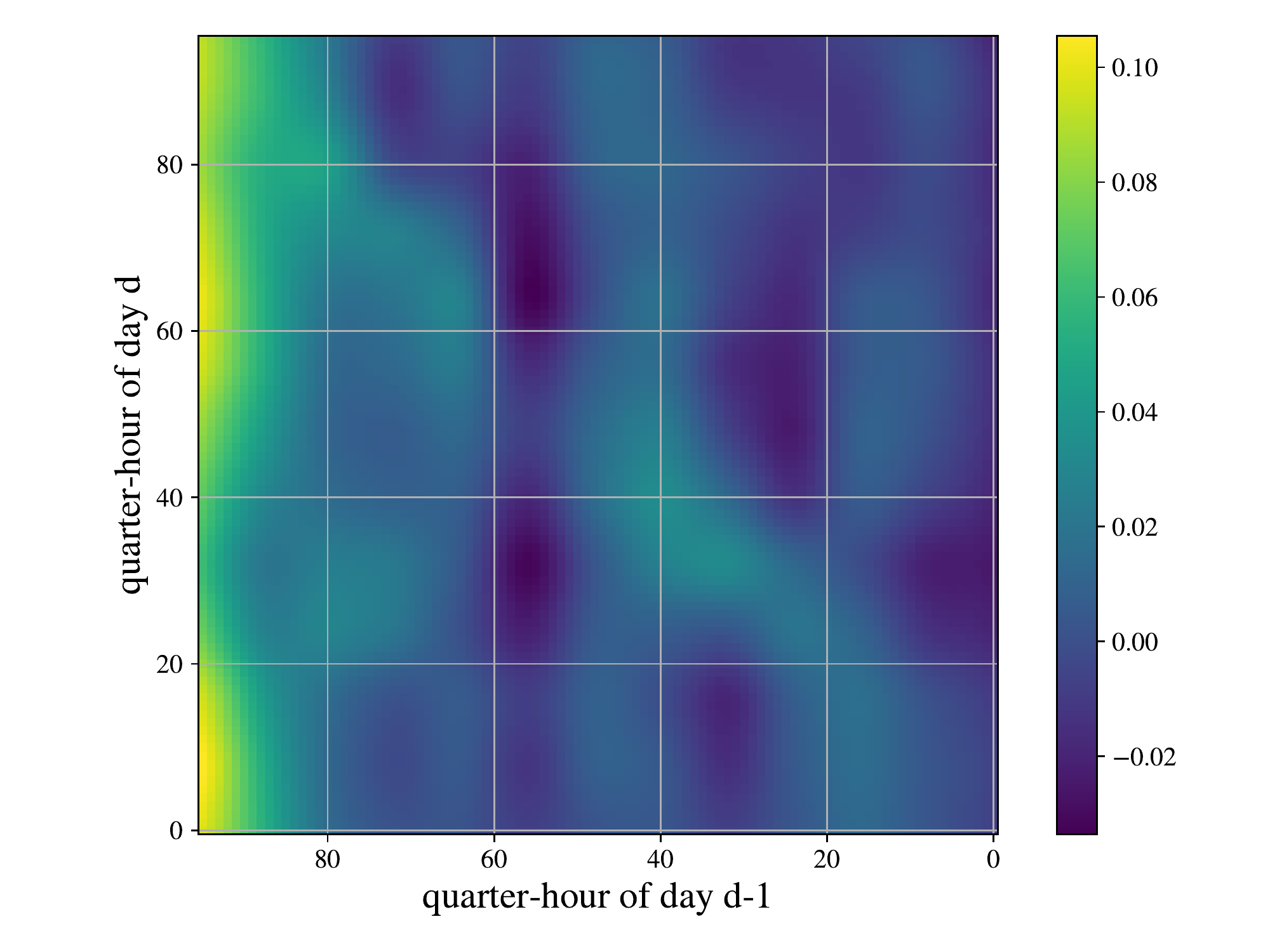} \label{fig:incre8bottomright}} 
\caption{Regularized Radial Basis Functions weight surface estimated on 2018 data: cubic polynomial component (top left), Gaussian component (top right), final Regularized Radial Basis Functions surface 3D view (bottom left) and top view (bottom right).}
\label{fig:incre8}
\end{figure*}

The cubic surface component $\mathbf{\bar{A}}$, the regularized radial basis function surface $\mathbf{\tilde{A}}$ and the final surface $\mathbf{A}$ are shown in Fig. \ref{fig:incre8topleft}, \ref{fig:incre8topright} and \ref{fig:incre8bottomleft}, respectively.

Again, the visual inspection of the Top view (Panel (d) of Fig. \ref{fig:incre8}) reveals the presence of two `edges', one horizontal and one diagonal, highlighted by the yellow/light green color. The dominance of such edges motivates the exploration of the sparse identification strategy, described in the next subsection.

\subsection{Two-edges model (TE)}
We now consider a simplified sparse model of the weight surface $\mathbf{A}$. The name `Two-edges' is due to the fact that only two vectors (herein called `edges') are estimated instead of a full $96\times96$ surface of weights, namely the last column
\begin{equation*}
\mathbf{a}^{last} =\left[\begin{array}{cccc} a_{1,96} & a_{2,96} & \ldots & a_{95,96} \end{array}\right]^{T} \in R^{95  \times 1}
\end{equation*}
and the main diagonal 
\begin{equation*}
\mathbf{a}^{diag} =\left[\begin{array}{cccc} a_{1,1} & a_{2,2} & \ldots & a_{96,96} \end{array}\right]^{T} \in R^{96  \times 1}
\end{equation*}
of $\mathbf{A}$, which leads to a model with $191$ parameters (note that $a_{96,96}$ is shared by the last column and the main diagonal).

The new model formulation is
\begin{equation*}
\mathbf{\hat{y}} = \mathbf{X} \bm{\beta}
\end{equation*}
with $\bm{\beta}\left(\mathbf{a}^{diag}, \mathbf{a}^{last}\right) = \left[\begin{array}{cccccc} a_{1,1} & a_{1,96} & a_{2,2} & a_{2,96} & \ldots & a_{96, 96}  \end{array}\right]^{T} \in R^{191  \times 1}$ and

\begin{equation*}
  \mathbf{\Psi}=
  \left[ {\begin{array}{c}
   \mathbf{I}_{96\times96} \otimes \boldsymbol{\psi}_{k}^{T}(0)  \\
    \mathbf{I}_{96\times96} \otimes \boldsymbol{\psi}_{k}^{T}(1) \\
    \vdots \\
    \mathbf{I}_{96\times96} \otimes \boldsymbol{\psi}_{k}^{T}(n_{day} - 1)  \\
  \end{array} } \right]^{T}
\in R^{96n_{day}  \times 1} 
\end{equation*}

with $\boldsymbol{\psi}_{k}(j) =\left[\begin{array}{cc} y(j, k) & y(j, 96) \end{array}\right]^{T}$.

Further regularization can be introduced by adding two penalty terms $\lambda_{diag}$ and $\lambda_{last}$ that shrink the second derivatives of each edge. The optimization problem becomes
\begin{equation*}
\bm{\beta} = \argmin_{\bm{\beta}}(\mathbf{y} - \mathbf{\Psi}\bm{\beta})^{T}(\mathbf{y} - \mathbf{\Psi}\bm{\beta}) + \lambda_{last} \Vert \mathbf{\Delta}_{last} \bm{\beta} \Vert_{2}^{2} + \lambda_{diag} \Vert \mathbf{\Delta}_{diag} \bm{\beta} \Vert_{2}^{2}
\end{equation*}
where $\mathbf{\Delta}_{last}\in R^{94 \times 191}$ and $\mathbf{\Delta}_{diag}\in R^{94 \times 191}$ are such that

\begin{equation*}
\mathbf{\Delta}_{last}\bm{\beta} = 
\begin{bmatrix} 
a_{3,96} - 2a_{2,96} + a_{1,96} \\
a_{4,96} -2 a_{3,96} + a_{2,96} \\
\vdots \\
a_{96,96} -2a_{95,96} + a_{94,96}
\end{bmatrix},
\end{equation*}

\begin{equation} \label{eq:delta_diag}
\mathbf{\Delta}_{diag}\bm{\beta} = 
\begin{bmatrix} 
a_{3,3} - 2a_{2,2} + a_{1,1} \\
a_{4,4} - 2a_{3,3} + a_{2,2} \\
\vdots \\
a_{96,96} - 2a_{95,95} + a_{94,94} 
\end{bmatrix}
\end{equation}
\\

The corresponding $\mathbf{T}$ matrix is
\begin{equation*}
\mathbf{T} = \lambda_{last} \mathbf{\Delta}^{T}_{last}\mathbf{\Delta}_{last} + \lambda_{diag}\mathbf{\Delta}^{T}_{diag}\mathbf{\Delta}_{diag}.
\end{equation*}

\begin{figure}
\centering
\includegraphics[width=.8\textwidth]{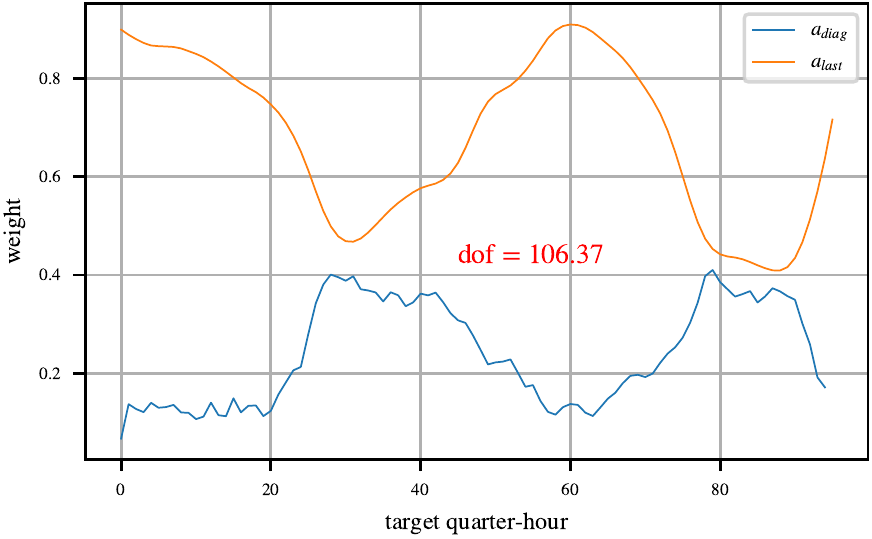}
\caption{Behaviour of the parameters of the TE model as a function of the quarter-hour of the target value.}
\label{fig:incre9}
\end{figure}

The estimated parameters $\mathbf{a}^{last}$ and $\mathbf{a}^{diag}$ are shown in Fig. \ref{fig:incre9}. It is clear that $\mathbf{a}^{diag}$ and $\mathbf{a}^{last}$ are correlated, which suggests a further simplification of the parametrization, which is discussed in the next subsection.

\subsection{One-edge model (OnE)}
The `One-edge' model reduces the weight matrix $\mathbf{A}$ just to its diagonal entries, that is the vector of parameters $a^{diag}$. According to this model, the 7-day difference at a certain quarter-hour of the target day is proportional to the 7-day difference of the previous day at the same quarter-hour. 

In the resulting model, $\bm{\beta} = \mathbf{a}_{diag}$ and $\mathbf{X} = \mathbf{\Xi}$ with

\begin{equation*}
  \mathbf{\Xi}=
  \left[ {\begin{array}{c}
   \diag\left(\mathbf{Y}(0)\right)  \\
    \diag\left(\mathbf{Y}(1)\right) \\
    \vdots \\
    \diag\left(\mathbf{Y}(n_{day} - 1)\right)  \\
  \end{array} } \right]^{T}
\in R^{96n_{day}  \times 96} \end{equation*}

where the diagonalization operator $\diag(\cdot)$ is defined as

\begin{equation*}
  \diag\left(\mathbf{Y}(d)\right) =
  \left[ {\begin{array}{ccccc}
   y_{d,1} & 0 & 0 & \ldots & 0 \\
0 & y_{d, 2} & 0 & \ldots & 0 \\
0 & 0 & y_{d, 3} & \ldots & 0 \\
\vdots & \vdots & \vdots & \ddots & \vdots \\
0 & 0 & 0 & 0 & y_{d, 96}
  \end{array} } \right]
\in R^{96 \times 96} 
\end{equation*}

A penalty term on the second derivative of the parameter vector is included in the cost function:
\begin{equation*}
\bm{\beta} = \argmin_{\bm{\beta}}(\mathbf{y} - \mathbf{\Xi}\bm{\beta})^{T}(\mathbf{y} - \mathbf{\Xi}\bm{\beta}) + \lambda_{diag} \Vert \mathbf{\Delta}_{diag} \bm{\beta} \Vert_{2}^{2}
\end{equation*}
with $\mathbf{\Delta}_{diag}$ defined as in \eqref{eq:delta_diag}. The corresponding $\mathbf{T}$ matrix is given by
\begin{equation*}
\mathbf{T} = \lambda_{diag}\mathbf{\Delta}^{T}_{diag}\mathbf{\Delta}_{diag}.
\end{equation*}

\begin{figure}
\centering
\includegraphics[width=.8\textwidth]{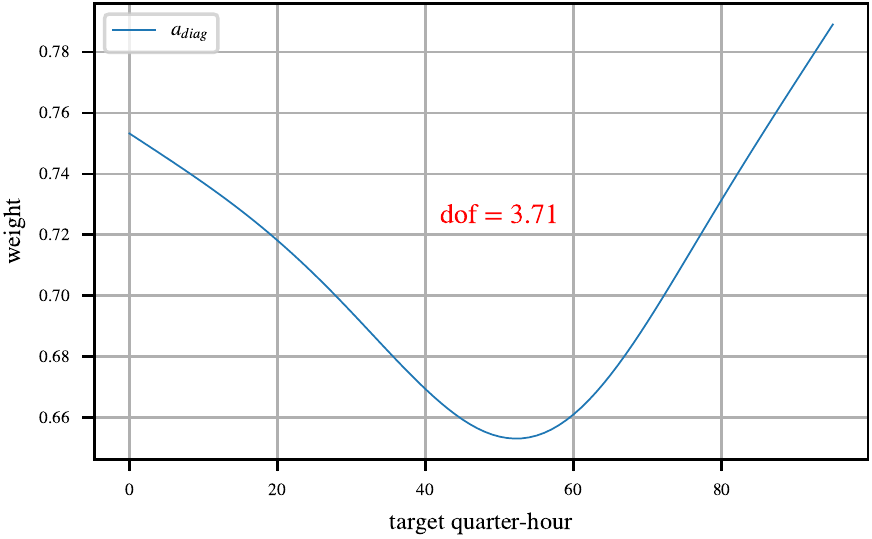}
\caption{One-edge model parameters behaviour as a function of the quarter-hour of the target value.}
\label{fig:incre10}
\end{figure}

The regularized solution, shown in Fig. \ref{fig:incre10}, is characterized by a dramatic decrease of the degrees of freedom, compared to the other models. Before illustrating the results of the prediction models, that will be presented in Section \ref{sec:Forecasting_results}, we introduce another methodology relying on an aggregation paradigm.

\subsection{Aggregated forecast} \label{sec:aggregated_forecast}
The combination of multiple forecasts is a commonly adopted technique in order to improve forecasts \cite{webb2004ensemblelearning}. A great variety of aggregation techniques has been proposed in the literature \cite{nowotarski2015sisterforecasts, wallis2011combiningforecasts, ranjan2010combiningprobabilityforecasts, devaine2013forecasting}. The simple average method is in general an effective and robust strategy adopted to reduce the prediction variance \cite{wallis2011combiningforecasts, ranjan2010combiningprobabilityforecasts, devaine2013forecasting}.

Let $\mathcal{M}_i, i=1,\ldots,m$, denote a set of prediction models and $\hat{L}_{\mathcal{M}_i}(d, q)$ the corresponding load demand forecasts generated by $\mathcal{M}_i$. Then
\begin{equation*}
\hat{L}_{\text{Avg}}(d, q) =\frac{1}{m} \sum_{i=1}^{m}\hat{L}_{\mathcal{M}_i}(d, q) 
\end{equation*}
is the aggregated forecast obtained by averaging the predictions generated by $\mathcal{M}_i$.

In order to better understand the potential benefit of aggregation, consider the simple case of just two prediction models, i.e. $m=2$, and
define the corresponding residuals as:
\begin{equation*}
e_{\mathcal{M}_i}(d, q) = L(d,q) - \hat{L}_{\mathcal{M}_i}(d, q)
\end{equation*}
It is immediate to verify that
\begin{equation*}
e_{\text{Avg}}(d, q) = L(d,q) - \hat{L}_{\text{Avg}}(d, q)=\frac{1}{2}\left(e_{\mathcal{M}_1}(d, q) + e_{\mathcal{M}_2}(d, q) \right)
\end{equation*}
Then the Mean Squared Error (MSE) of the aggregated forecast is given by:
\begin{equation}
\label{eq:mse_equation}
\begin{split}
MSE_{\text{Avg}} &= E\left[e_{\text{Avg}}^{2}\right] = E\left[\frac{1}{4}\left(e_{\mathcal{M}_1}^2 + e_{\mathcal{M}_2}^2 + 2e_{\mathcal{M}_1}e_{\mathcal{M}_2} \right)\right] = \\ &= \frac{1}{4}\left( E\left[e_{\mathcal{M}_1}^2\right] + E\left[e_{\mathcal{M}_2}^2\right] + 2E\left[e_{\mathcal{M}_1}e_{\mathcal{M}_2}\right]\right) = \\
&= \frac{1}{4}MSE_{\mathcal{M}_1} + \frac{1}{4}MSE_{\mathcal{M}_2} + \\ &+ \frac{1}{2}\left(Cov\left[e_{\mathcal{M}_1}e_{\mathcal{M}_2}\right] + E\left[e_{\mathcal{M}_1}\right]E\left[e_{\mathcal{M}_2}\right]\right) = \\
&= \frac{1}{4}MSE_{\mathcal{M}_1} + \frac{1}{4}MSE_{\mathcal{M}_2} + \\ &+ \frac{1}{2}\left(\rho_{e_{\mathcal{M}_1}, e_{\mathcal{M}_2}}\sigma_{e_{\mathcal{M}_1}}\sigma_{e_{\mathcal{M}_2}} + E\left[e_{\mathcal{M}_1}\right]E\left[e_{\mathcal{M}_2}\right] \right)
\end{split}
\end{equation}
where $\rho_{e_{\mathcal{M}_1}, e_{\mathcal{M}_2}}$ is the coefficient of correlation between $e_{\mathcal{M}_1}$ and $e_{\mathcal{M}_2}$.

Assume that $\mathcal{M}_1$ performs better than $\mathcal{M}_2$, that is $MSE_{\mathcal{M}_1} < MSE_{\mathcal{M}_2}$. Then the aggregated forecaster $\hat{L}_{\text{Avg}}(d, q)$ improves on $\hat{L}_{\mathcal{M}_1}(d, q)$ provided that the following inequality is satisfied:
\begin{equation*}
MSE_{\mathcal{M}_2} < 3MSE_{\mathcal{M}_1} - 2E[e_{\mathcal{M}_1} e_{\mathcal{M}_2}] - 2 b_1 b_2
\end{equation*}
where $bias_{1} = E\left[e_{\mathcal{M}_1}\right]$ and $bias_{2} = E\left[e_{\mathcal{M}_2}\right]$ are the bias of $\mathcal{M}_{1}$ and $\mathcal{M}_{2}$ respectively.

Consider for instance the case when at least one of the predictors has a negligible bias. Then, provided that the covariance between the errors $e_{\mathcal{M}_1}$ and $e_{\mathcal{M}_2}$ is small, the aggregation between the two predictors brings an improvement even when the MSE of the worse one is up to three times larger than that of the best one. As a consequence, there is room for designing predictors that employ different strategies aiming at obtaining scarcely correlated prediction errors. Later on, the correlation between the residuals of the newly proposed method and those of the Italian TSO predictor will be studied in order to assess the potential benefits ensuing from an aggregation.


\section{Experimental validation setup} \label{sec:Experimental_validation_setup}

Three scenarios were considered for evaluating and comparing the proposed models:
\begin{itemize}
\item[-] Training = 2016, Test = 2017
\item[-] Training = 2017, Test = 2018
\item[-] Training = 2018, Test = 2019
\end{itemize}
This choice is driven by the fact that the Terna forecasts, that are used as benchmark for comparison purposes, are available for the years 2017, 2018, 2019. 

The performances of the models were evaluated using three metrics: Mean Absolute Percentage Error ($MAPE$), Root Mean Square Error ($RMSE$), and Mean Absolute Error ($MAE$), each one both on quarter-hourly and daily data, for a total of six performance indexes:
\begin{equation*}
	MAPE = \frac{100}{n} \sum_{d \in \mathpzc{D_{Te}}} \sum_{q=1}^{96} \left| \frac{L(d, q) -  \hat{L}(d, q)}{L(d,q)} \right| 
\end{equation*}
\begin{equation*}
	RMSE = \sqrt{\frac{\sum_{d \in \mathpzc{D_{Te}}} \sum_{q=1}^{96} \left(L(d,q) - \hat{L}(d,q) \right)^{2}}{n}} 
\end{equation*}
\begin{equation*}
	MAE = \frac{\sum_{d \in \mathpzc{D_{Te}}} \sum_{q=1}^{96} \left| L(d, q) -  \hat{L}(d, q) \right|}{n} 
\end{equation*}
\begin{equation*}
	MAPE_{daily} = \frac{100}{n_{day}} \sum_{d \in \mathpzc{D_{Te}}} \left| \frac{L_{daily}(d) -  \hat{L}_{daily}(d)}{L_{daily}(d)} \right| 
\end{equation*}
\begin{equation*}
	RMSE_{daily} = \sqrt{\frac{\sum_{d \in \mathpzc{D_{Te}}} \left( L_{daily}(d) - \hat{L}_{daily}(d) \right)^{2}}{n_{day}}} 
\end{equation*}
\begin{equation*}
	MAE_{daily} = \frac{\sum_{d \in \mathpzc{D_{Te}}} \left|  L_{daily}(d) -  \hat{L}_{daily}(d) \right|}{n_{day}}
\end{equation*}

where $\mathpzc{D_{Te}}$ is the set of test days, $n$ is the number of quarter-hourly test data, and $n_{day}$ is the number of daily test data. In particular, $\mathpzc{D_{Te}}=\{d: d \not\in \mathpzc{D}_{s}, d-7 \not\in \mathpzc{D}_{s}\}$, where $\mathpzc{D}_{s}$ includes special days such as Winter, Summer, Easter and national holidays (see Appendix). $L_{daily}$ and $\hat{L}_{daily}$ are respectively the time series of daily averages of electric load observations and the associated forecasts:
\begin{eqnarray*}
L_{daily}(d) = \frac{1}{96}\sum_{q=1}^{96}L(d, q), \quad
\hat{L}_{daily}(d) = \frac{1}{96}\sum_{q=1}^{96}\hat{L}(d, q)
\end{eqnarray*}

The hyperparameters of the models described in Section \ref{sec:Forecasting_methods} are tuned through cross-validation choosing the $MAPE$ as objective function and using, for each scenario, the two years preceding the test one as training and validation: e.g. for the first scenario (test year = 2017) cross-validation is performed on the years 2015 and 2016, using the first year for training and the second one for validation.
All the results of the hyperparameters tuning phase are summarized in Table ~\ref{table:Table1}.

\input{Table1.tex}


\section{Forecasting results} \label{sec:Forecasting_results}

In this section, the predictive performances of the models described in Section \ref{sec:Forecasting_methods} are discussed and compared to the Terna forecaster. The results for the three test scenarios are summarized in Table ~\ref{table:Table2}, ~\ref{table:Table3}, ~\ref{table:Table4}, where, for each predictor, the performance indexes introduced in Section \ref{sec:Experimental_validation_setup} and the Degrees of Freedom ($\mathrm{dof}$), accounting for the complexity of the underlying models, are reported. 

\input{Table2.tex}

\input{Table3.tex}

\input{Table4.tex}

\begin{figure*}[!htb]
  \centering 
  \subfigure{\includegraphics[scale=0.67]{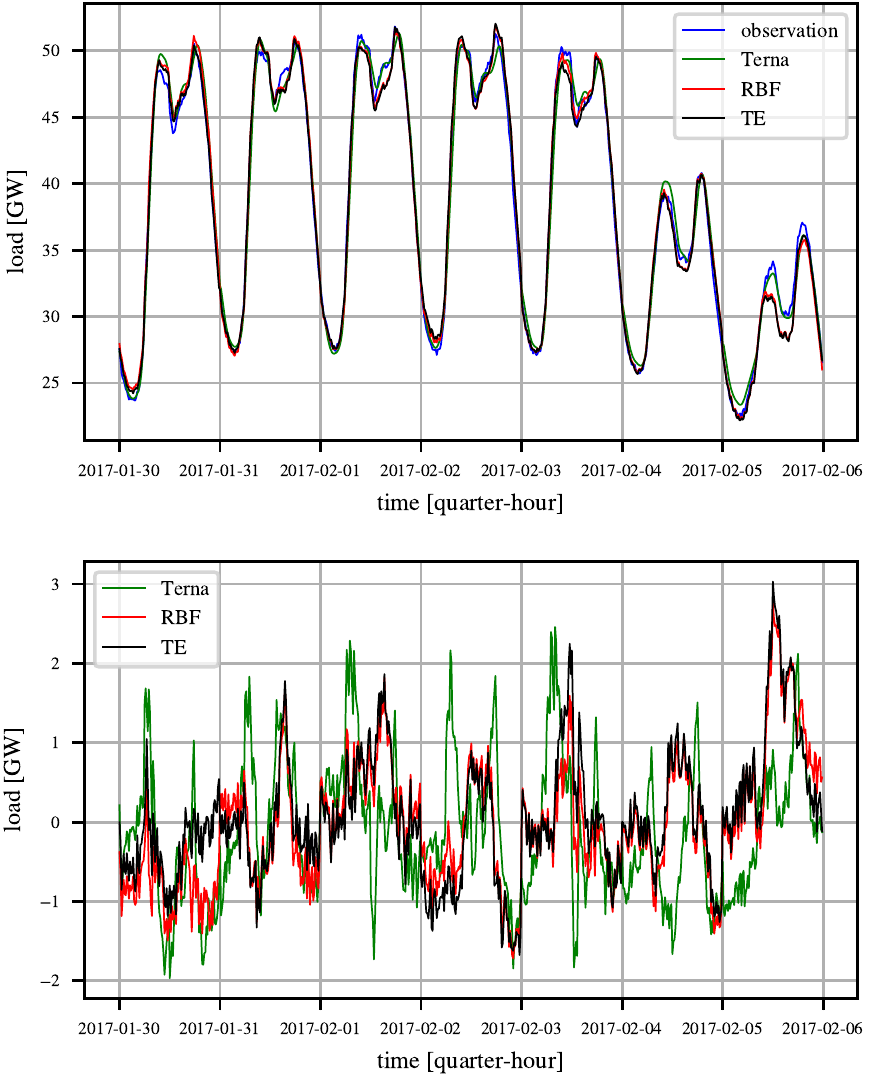}}
  \subfigure{\includegraphics[scale=0.67]{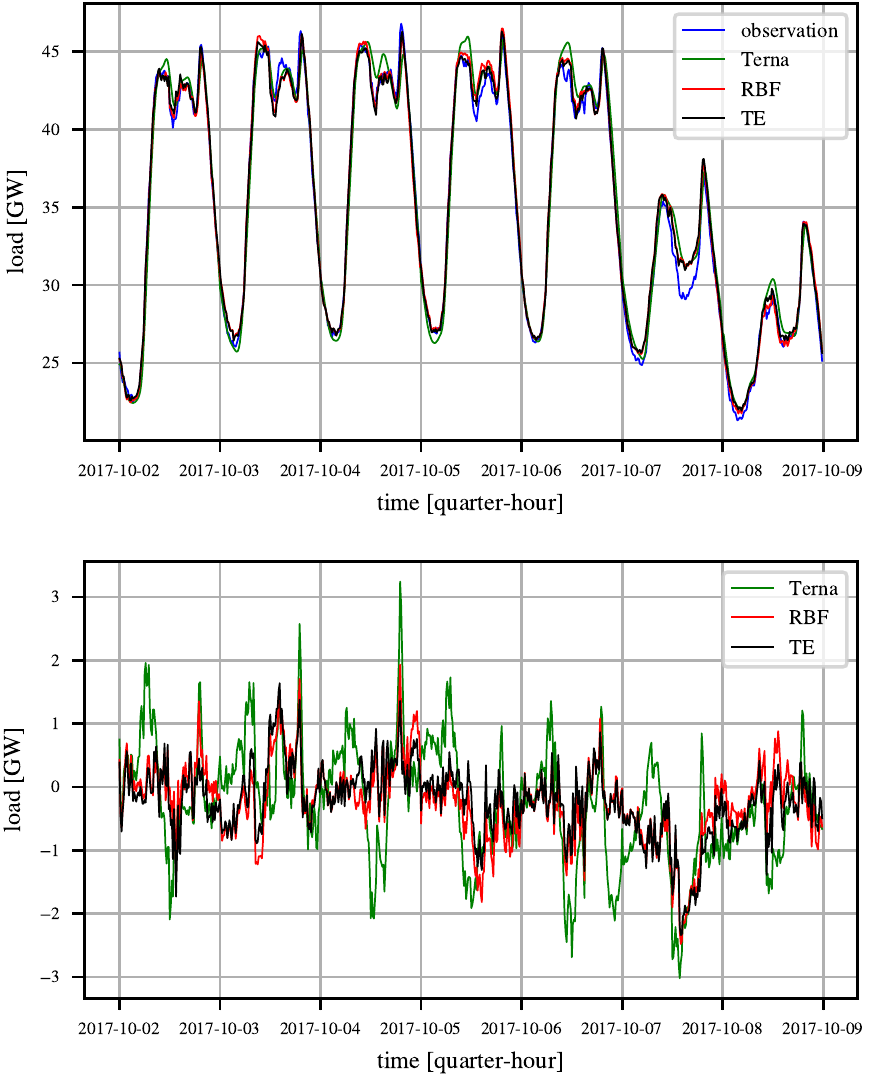}} 
\caption{RBF, TE and Terna prediction (top) and residual (bottom) over two sample weeks on 2017.}
\label{fig:incre11}
\end{figure*}

\begin{figure*}[!htb]
  \centering 
  \subfigure{\includegraphics[scale=0.67]{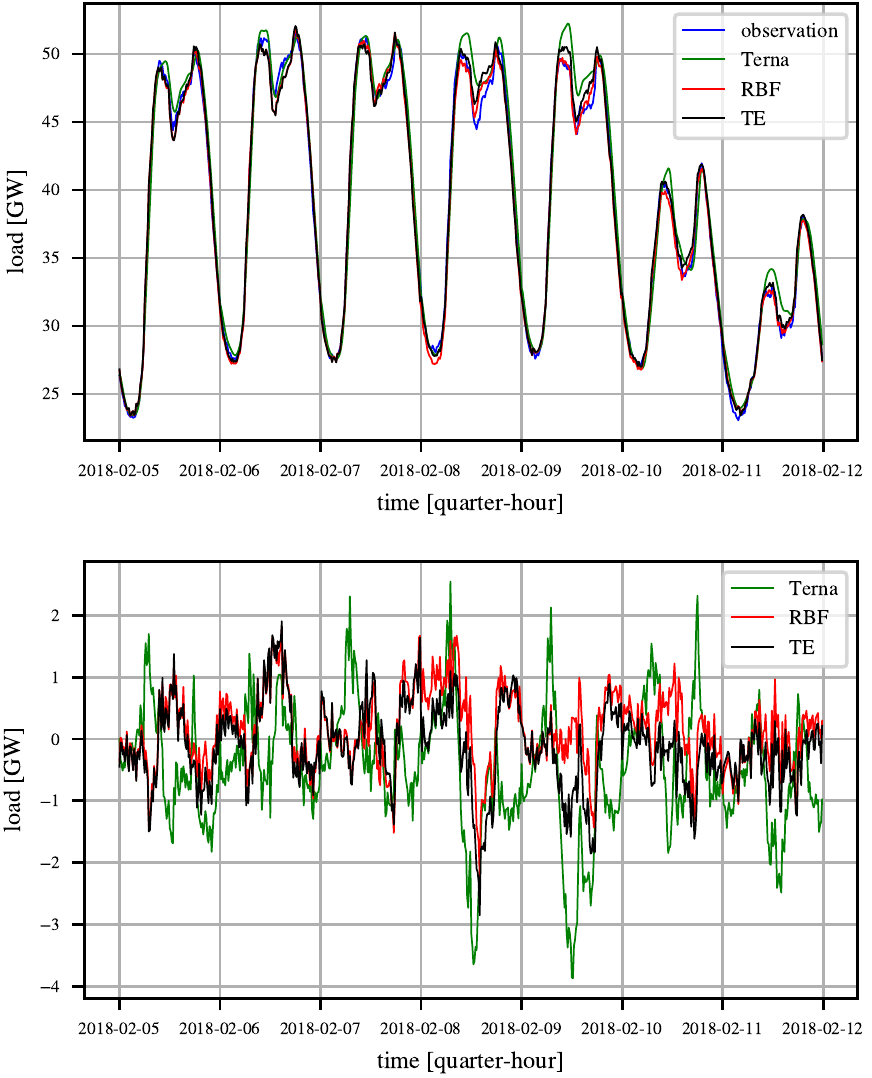}} 
  \subfigure{\includegraphics[scale=0.67]{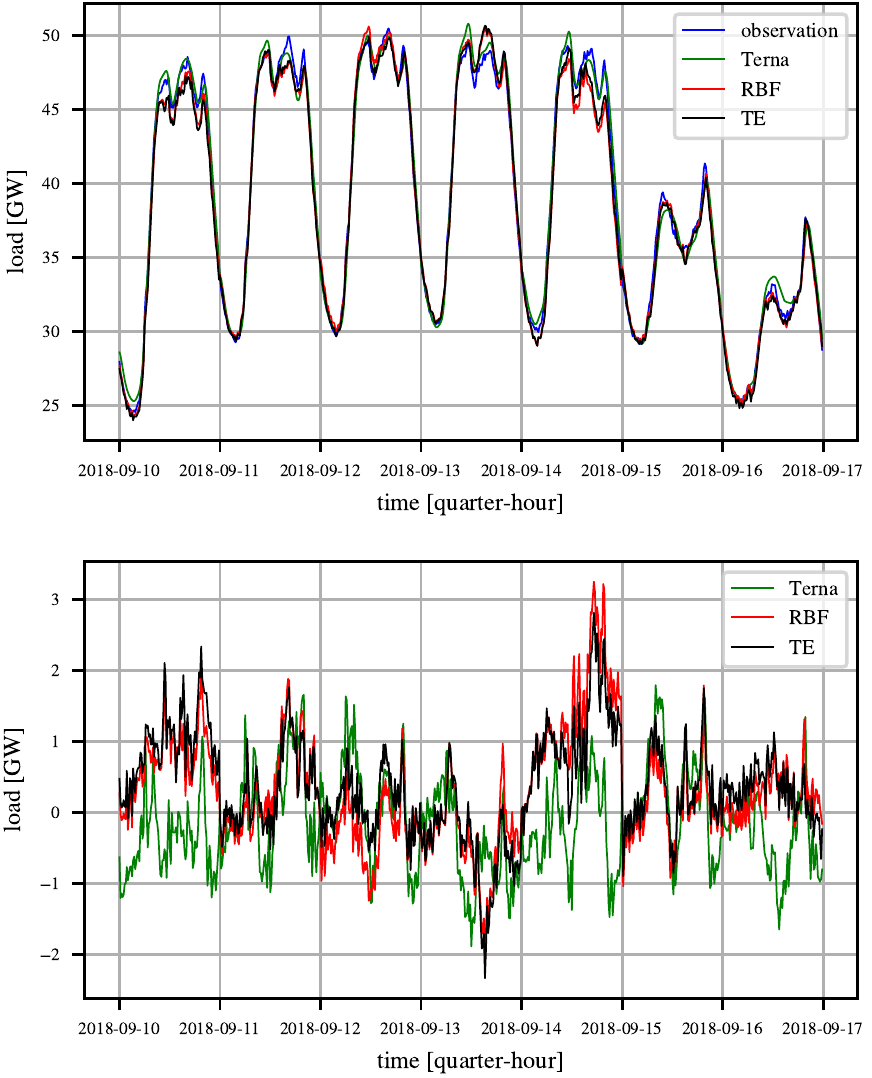}} 
\caption{RBF, TE and Terna prediction (top) and residual (bottom) over two sample weeks on 2018.}
\label{fig:incre12}
\end{figure*}

\begin{figure*}[!htb]
  \centering 
  \subfigure{\includegraphics[scale=0.67]{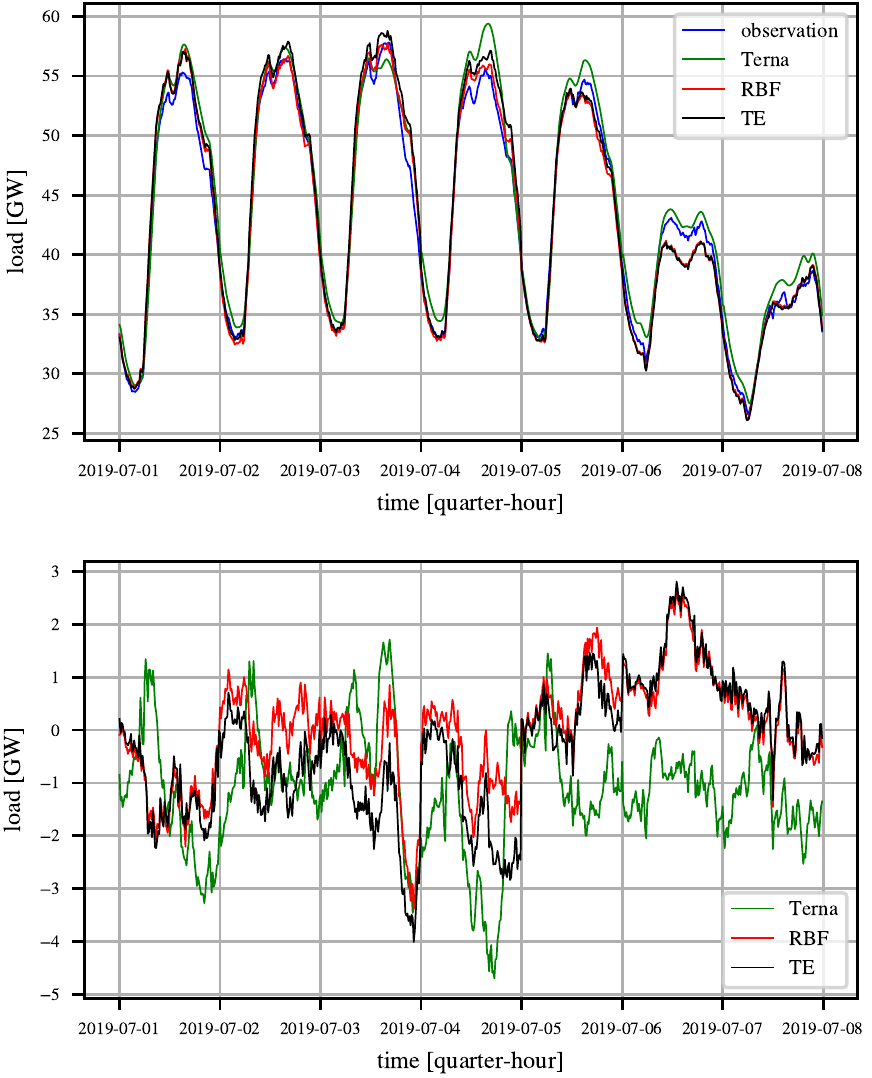}} 
  \subfigure{\includegraphics[scale=0.67]{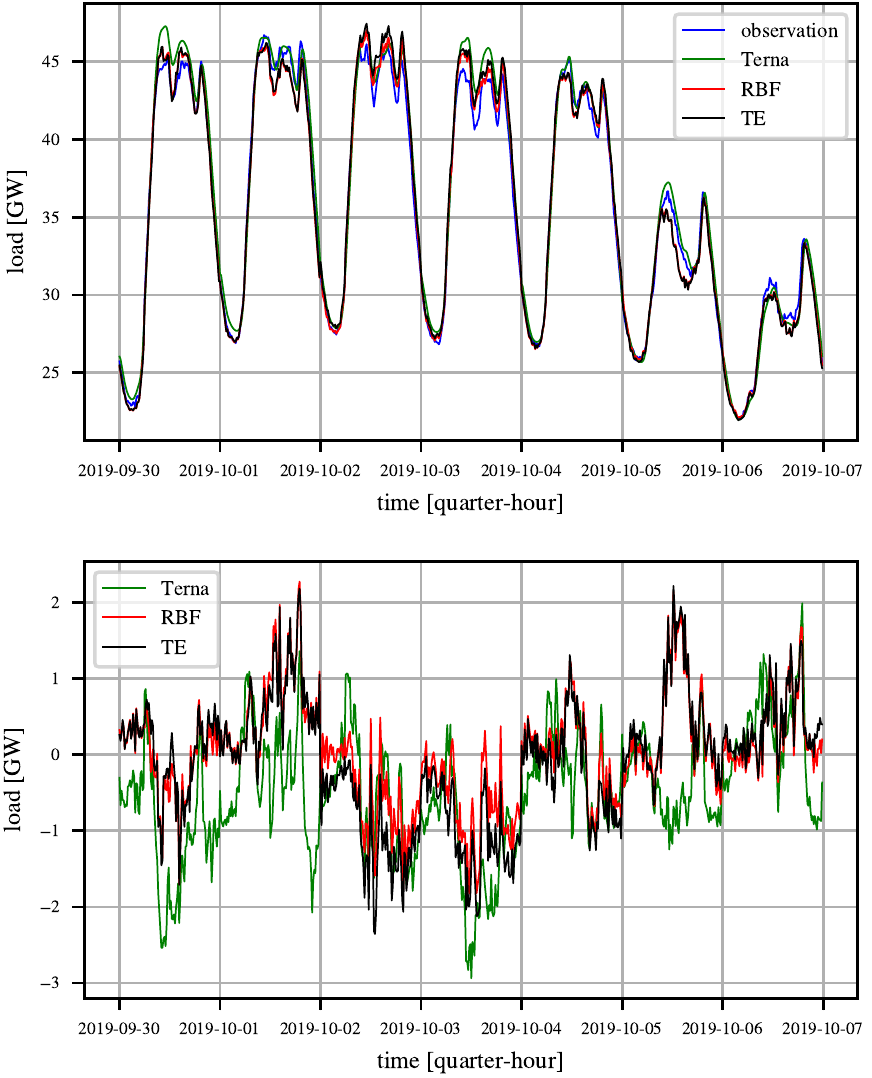}} 
\caption{RBF, TE and Terna prediction (top) and residual (bottom) over two sample weeks on 2019.}
\label{fig:incre13}
\end{figure*}

\begin{figure*}[!htb]
  \centering 
  \subfigure{\includegraphics[scale=0.16]{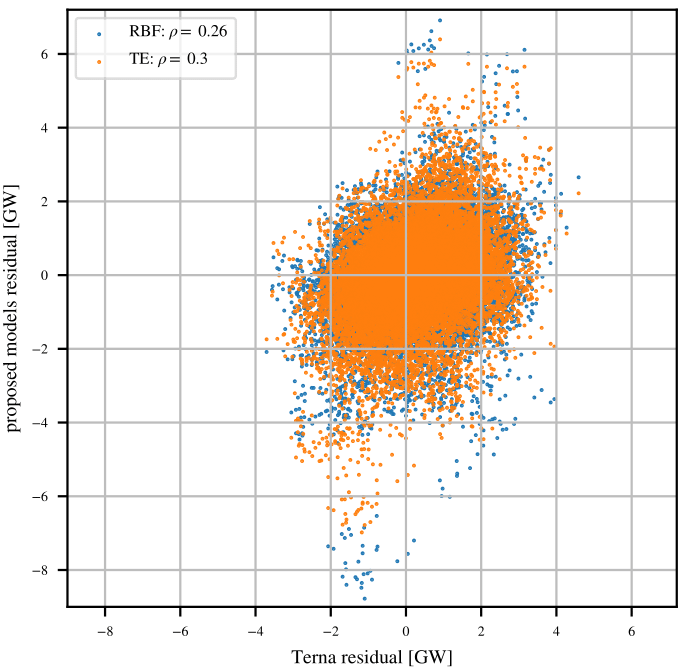}}
  \subfigure{\includegraphics[scale=0.16]{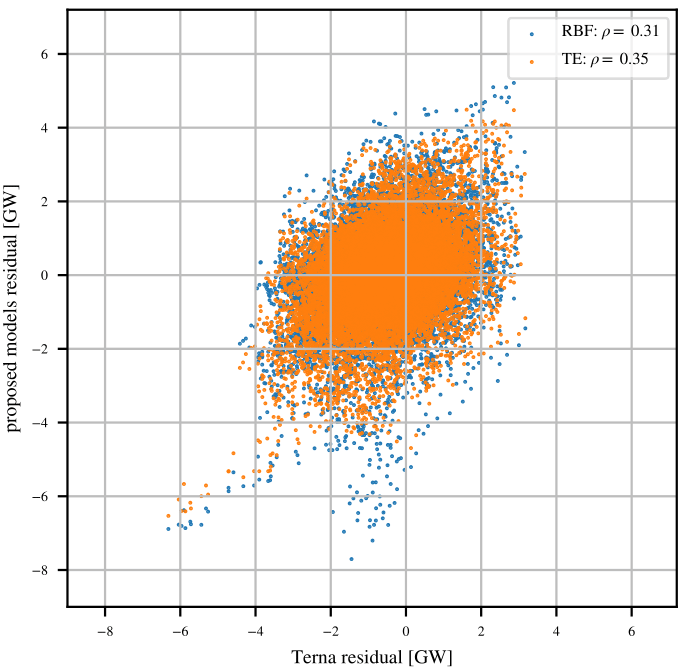}}
  \subfigure{\includegraphics[scale=0.16]{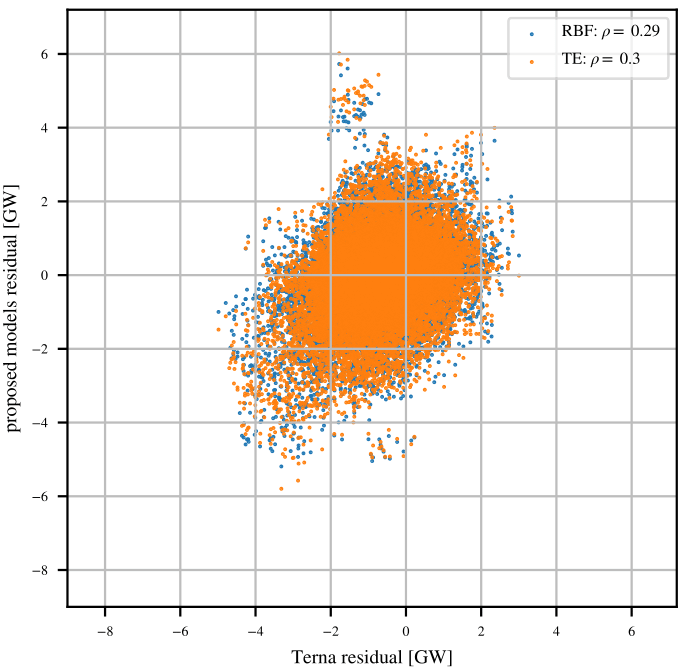}}
\caption{Scatter plots between the Terna residual and RBF (blue dots) and TE (orange dots) model residuals on the three test scenarios (left: 2017, middle: 2018, right: 2019). The Pearson correlation coefficients indicate a weak correlation between the proposed models errors and the Terna forecast error.}
\label{fig:incre14}
\end{figure*}

In all scenarios the OLS approach performs poorly: it achieves the worst performances and has too many degrees of freedom (96 x 96 = 9216 = \# of parameters). On the other hand, the OnE model is too parsimonious: while it achieves results that improve on the OLS approach and in some cases are comparable to the benchmark ones (e.g. see 2018 and 2019 $MAPE$), its performances are significantly worse on the 2017 scenario and in all daily indexes. Significant improvements are obtained by resorting to regularized predictors. The TA,  TS and RBF models achieve comparable results over the three test years,  the main difference being represented by the complexity of the three approaches, where the last predictor stands out for being the most parsimonious one ($\mathrm{dof}$ around 178 in the three scenarios). 

The TE model, which further reduces the complexity of the predictor ($\mathrm{dof}$ in the range from 100 to 115) ranks first in the 2018 scenario, while it is slightly inferior to the RBF predictor in the other two scenarios. In view of this, it provides an effective compromise between accuracy and simplicity.

The percentage decreases of $MAPE$ and the $MAE$ brought by TA, TS, RBF and TE with respect to the benchmark  reach 20\% in 2018 and 24\% in 2019, while they are less evident in terms of $RMSE$ (2\% in 2018, 12\% in 2019). This can be explained by a better accuracy of the proposed predictors during the night (where there are lower demands) and by the presence of few large errors within the predictions (to which the $RMSE$ score is more sensitive than $MAPE$ and $MAE$).

The 2017 scenario is the tougher one. In particular, while percentage improvements on the quarter-hourly $MAPE$ and $MAE$ are relatively small (7\%), the quarter-hourly $RMSE$ results achieved by Terna are slightly better than ours. Moreover, the daily performances of Terna are superior than the ones achieved by the proposed models. 

This is the consequence of a few large errors within the proposed forecasts, possibly related to the fact that the proposed predictors do not account for any exogenous variables such as the temperature, which can be crucial for the prediction in some seasons of the year and some phases of the day. By contrast, this information is exploited by the Terna forecaster. In Fig. \ref{fig:incre11}, \ref{fig:incre12} and \ref{fig:incre13} it is possible to visualize how the forecasts of the RBF and TE models compare to the Terna ones over different weeks of 2017, 2018 and 2019. 

In particular, the residual plots reveal that, while the error profiles of the proposed models are similar, they are almost uncorrelated with the Terna forecast error, as confirmed by the inspection of the scatter plots in 2017, 2018 and 2019 of the residuals for Terna vs RBF (correlation coefficient $\rho_{e_{\mathcal{M}_{RBF}}, e_{\mathcal{M}_{Terna}}}$ ranging from 0.26 to 0.31) and Terna vs TE ($\rho_{e_{\mathcal{M}_{TE}}, e_{\mathcal{M}_{Terna}}}$ ranging 0.3 from to 0.35), see Fig. \ref{fig:incre14}. It turns out that the absolute values of the biases of RBF and TE are always less than 0.1. 

In view of this, there is room for improving the quality of the forecasts by combining Terna's predictions with those produced by the new proposed methods \cite{wallis2011combiningforecasts, devaine2013forecasting}. The margin for improvement was assessed by plugging into formula \eqref{eq:mse_equation} the values reported in Table \ref{table:Table5}. The formula predicts that a significant improvement can be achieved. 

In particular, the best $MSE$ that ranges from 0.98 (RBF) in 2019 to 1.13 $\left[GW^2\right]$  (Terna) in 2017 is predicted to range between 0.71 and 0.81 throughout the considered years when the aggregated predictors AVG(RBF) and AVG(TE) are employed. All the aggregated predictions obtained from the models proposed in Section \ref{sec:Forecasting_methods} were evaluated according to the same framework. The results, summarized in Tables ~\ref{table:Table6}, ~\ref{table:Table7}, ~\ref{table:Table8}, highlight a very substantial improvement with respect to Terna benchmark in all three scenarios, for all performances indexes and in both the quarter-hourly and the daily cases. In particular, the improvement with respect to the TSO benchmark for the quarter-hourly indexes is always not less than 20\%, reaching 32\% for the $MAPE$ 2019 (Avg(RBF)), while the improvement of the daily indexes is always not less than 15\%, reaching 35\% for $MAPE_{daily}$ and $MAE_{daily}$ in 2019 (Avg(RBF)). The predicted performances of the aggregated forecasters provided by formula (5.15) are remarkably accurate: the error is always not greater than 0.02 $\left[GW^2\right]$, see Table \ref{table:Table5}.

\begin{figure*}[!htb]
  \centering 
  \subfigure{\includegraphics[scale=0.67]{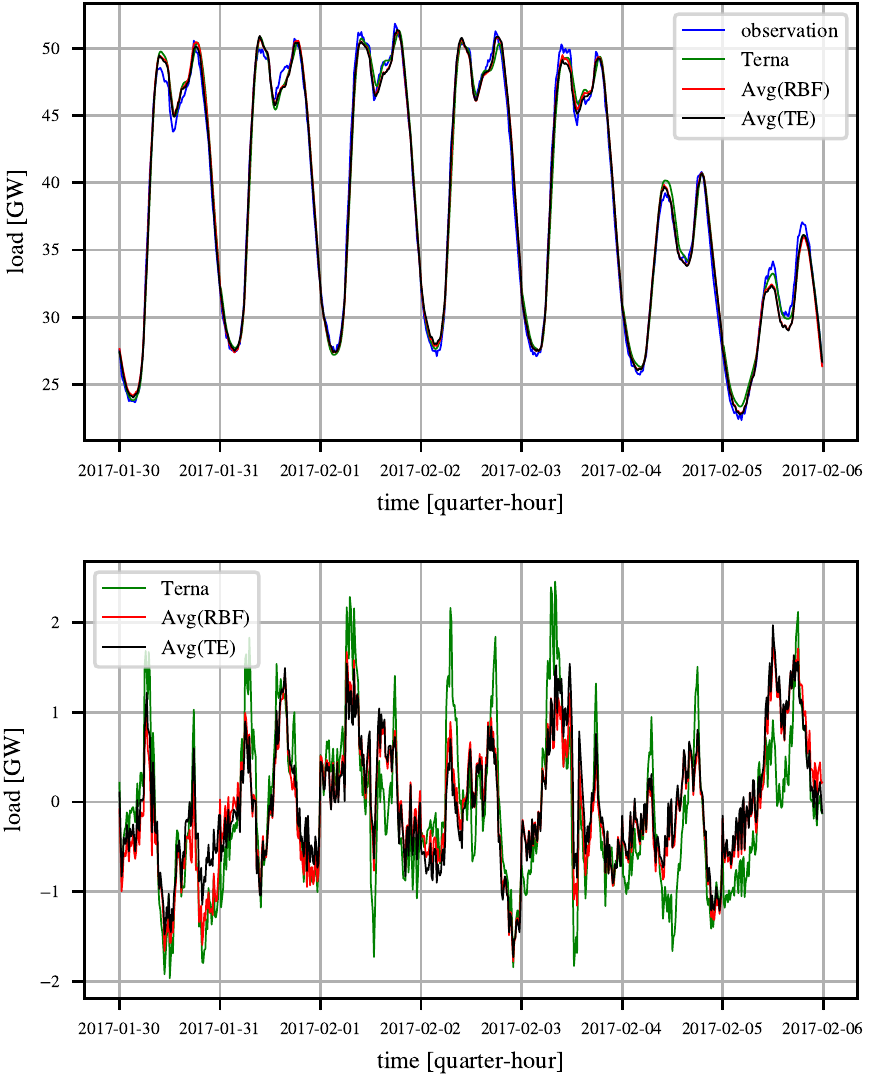}}
  \subfigure{\includegraphics[scale=0.67]{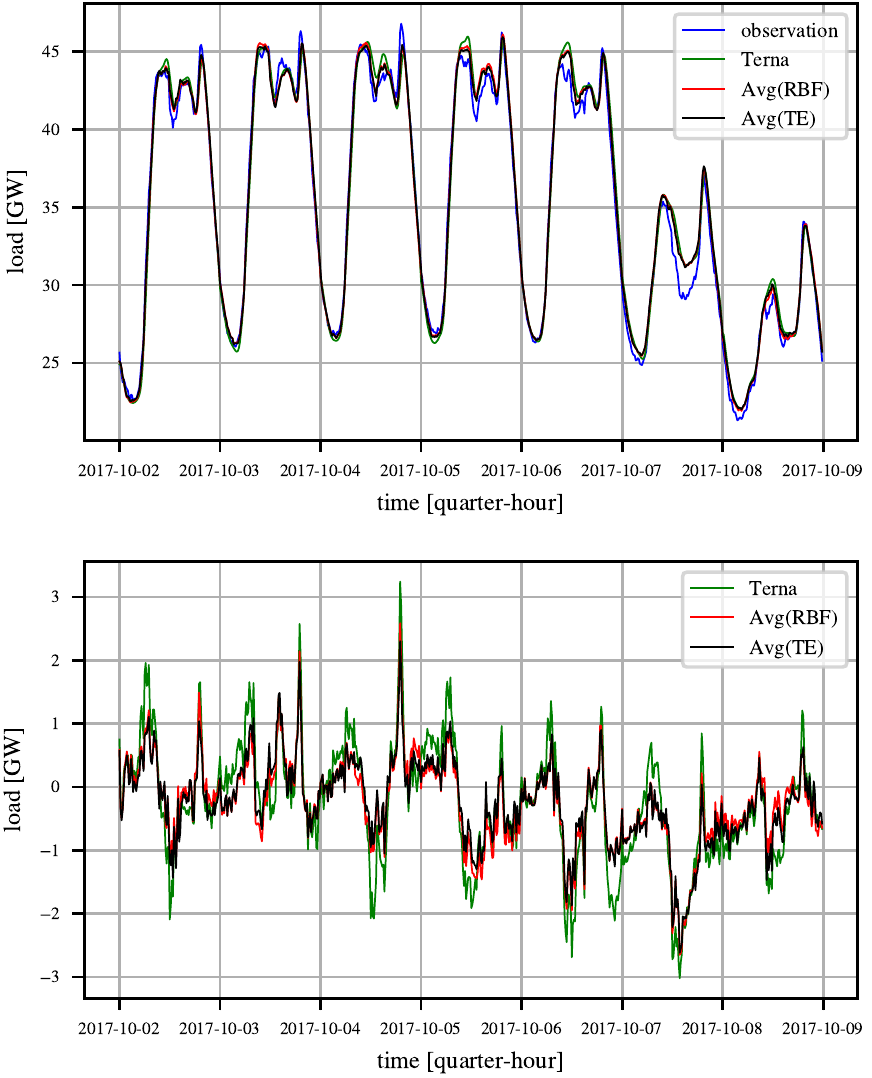}} 
\caption{Avg(RBF), Avg(TE) and Terna prediction (top) and residual (bottom) over two sample weeks on 2017.}
\label{fig:incre15}
\end{figure*}

\begin{figure*}[!htb]
  \centering 
  \subfigure{\includegraphics[scale=0.67]{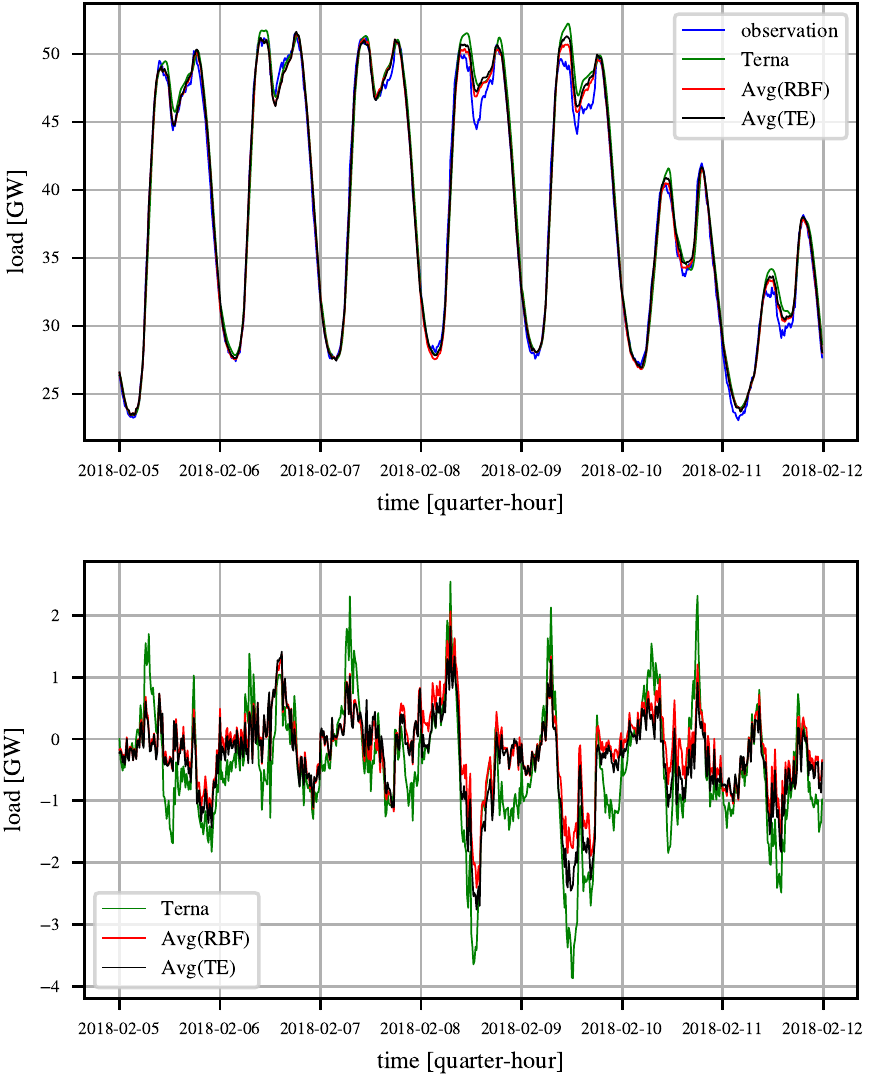}} 
  \subfigure{\includegraphics[scale=0.67]{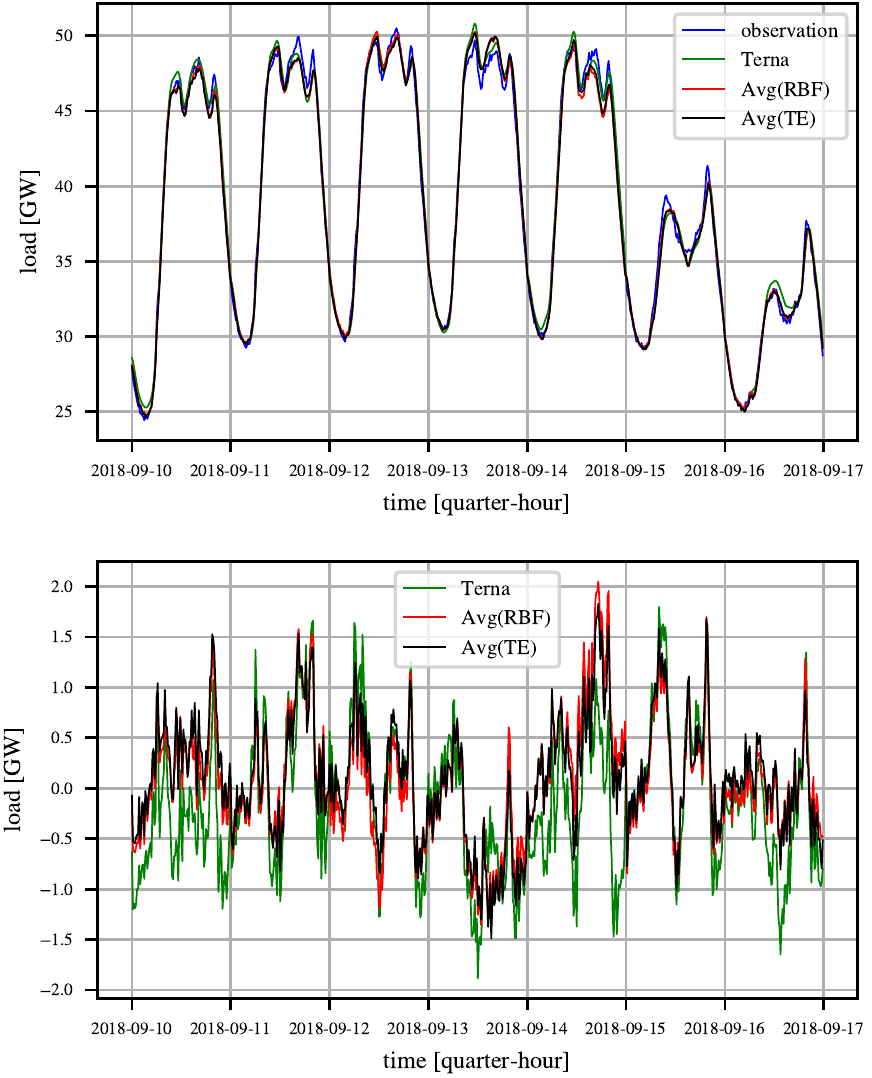}} 
\caption{Avg(RBF), Avg(TE) and Terna prediction (top) and residual (bottom) over two sample weeks on 2018.}
\label{fig:incre16}
\end{figure*}

\begin{figure*}[!htb]
  \centering 
  \subfigure{\includegraphics[scale=0.67]{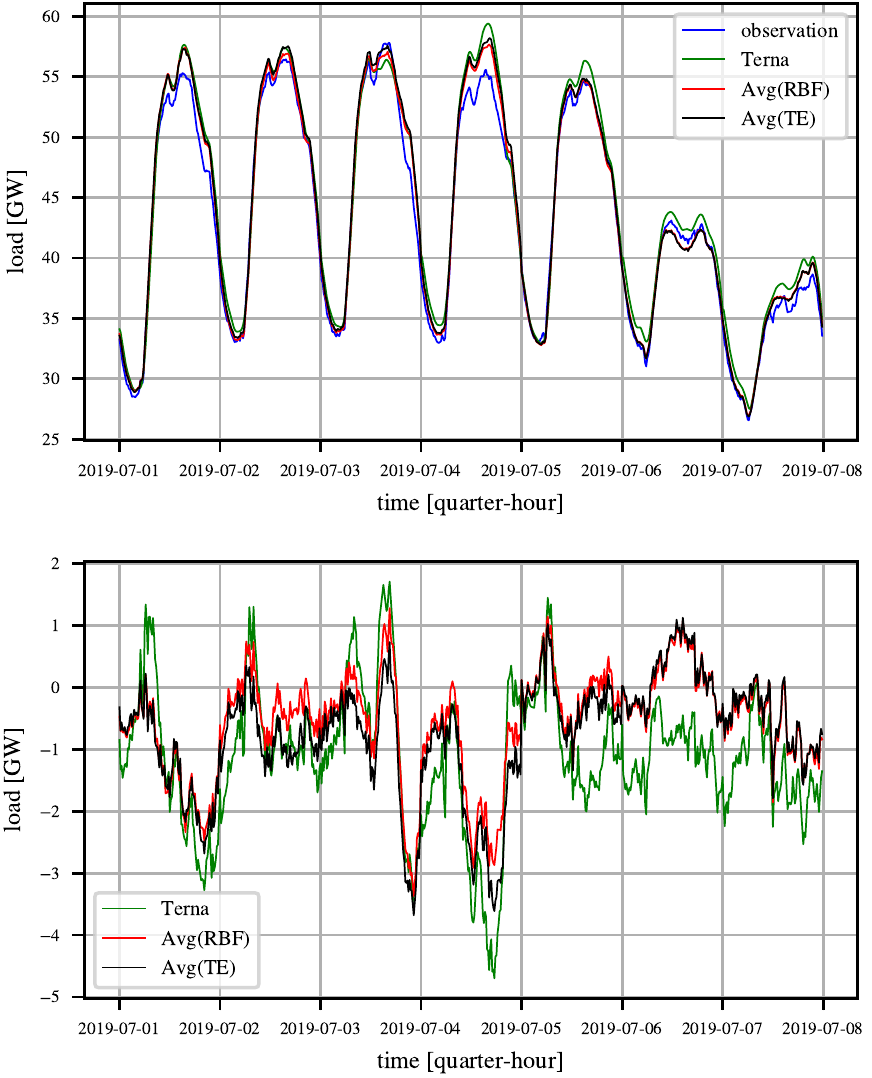}} 
  \subfigure{\includegraphics[scale=0.67]{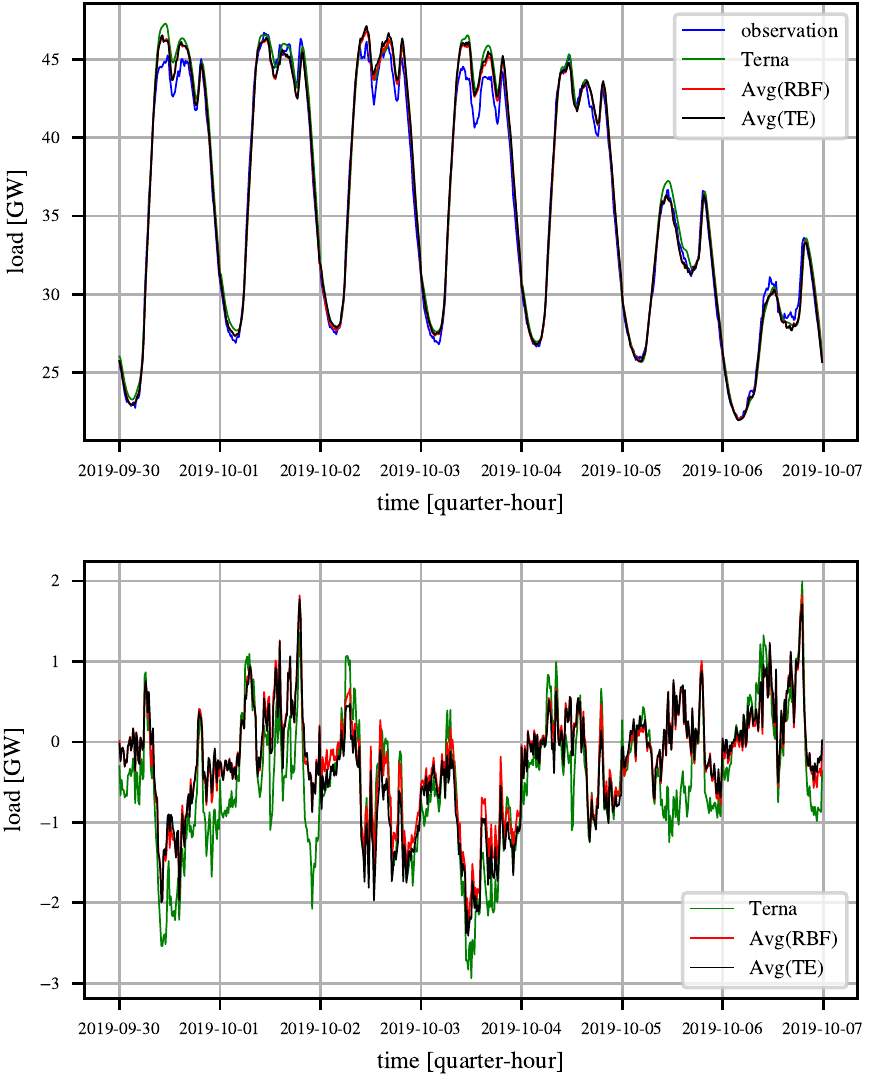}} 
\caption{Avg(RBF), Avg(TE) and Terna prediction (top) and residual (bottom) over two sample weeks on 2019.}
\label{fig:incre17}
\end{figure*}

The time plots and residual plots for the Avg(RBF) and the Avg(TE) cases over some sample weeks of 2017,2018 and 2019 are displayed in Fig. \ref{fig:incre15}, \ref{fig:incre16} and \ref{fig:incre17}. Overall, the new prediction strategy offers the opportunity for a significant reduction of the prediction errors, especially if considering an aggregated predictor that takes advantage of the uncorrelatedness of the errors committed by the new predictors and the Terna one.

\input{Table5.tex}

\input{Table6.tex}

\input{Table7.tex}

\input{Table8.tex}
\clearpage

\section{Conclusions} \label{sec:Conclusions}
We have shown that accurate one-day ahead predictions of the Italian electric load demand can be achieved on normal days, by a short-term predictor that does not model yearly seasonality and does not use exogenous information such as the one-day ahead prediction of the temperature. The considered prediction problem consists of predicting tomorrow's quarter-hourly demand profile based on the knowledge of today's profile until midnight. In particular we focused on the development of effective algorithms capable of exploiting the highly correlated nature of the signal. The first steps are a logarithmic transformation to achieve a more stable and symmetric signal and a 7-day differentiation that removes the weekly periodicity. The key idea behind the proposed forecaster is a multipredictor strategy, i.e. developing 96 linear predictors, each of which provides the prediction of the target signal during one of tomorrow's quarter-hours. In other words, each prediction is a linear combination of today's 96 samples. The full model, characterized by $96 \times 96 = 9216$ parameters, is obviously overparametrized so that different regularization approaches were employed to reduce the degrees of freedom without penalizing the predictive capabilities. The main observation is that the $9216$ parameters can be represented as a surface that, in view of the regularity of the load signal, should exhibit some smoothness properties.

The test results over 2017-2019 have shown that, through a wise application of regularization techniques it is possible to obtain competitive predictors whose MAPEs improve on that of
Terna. Moreover, the residuals of the proposed predictors are weakly correlated with Terna's, suggesting that aggregated forecasters could further improve the final results. 

As a matter of fact, averaging the Terna predictions with the proposed forecasts allows to reach an improvement up to the $30\%$ with respect to the Terna benchmark forecast in all performance indexes, both in a quarter-hourly and daily framework.

\section*{Funding}
This work has been partially supported by the Italian Ministry for Research in the framework of the 2017 Program for Research Projects of National Interest (PRIN), Grant no. 2017YKXYXJ.

\bibliography{mybibfile}

\section*{Appendix} \label{appendix:appendixA} 

The set $\mathpzc{D}_{s}$ of special days includes Winter, Summer, Easter and national holidays and their associated windows of influence. The days included in this set are summarized below.

\subsection*{\normalfont{\textit{Winter holidays}}}
Winter holidays account for Christmas Eve, Christmas, St. Stephen's Day, New Year's Eve, New Year and Epiphany holidays and include all days within December 22 and January 6, for a total of 16 days.

\subsection*{\normalfont{\textit{Summer holidays}}}
Summer holidays consists of a range of three weeks around August 15, in particular from August 5 to August 24 (20 days).

\subsection*{\normalfont{\textit{National holidays}}}
The national holidays in Italy are: Liberation Day (April 25), Labour Day (May 1), Republic Day (June 2), All Saints' Day (November 1), Feast of the Immaculate Conception (December 8). 
For each national holiday, a window of influence of five days is considered (two days before and after the holiday itself), for a total of 25 days.

\subsection*{\normalfont{\textit{Easter holidays}}}
Easter represent a particular case of holiday since its date is not fixed but it varies within March and April, while its weekday is always Sunday. In this work, a window of 5 days is associated to Easter holidays, in particular from the Thursday before to the Monday after (Easter Monday). According to this convention, the following table summarizes the dates of Easter holidays of the years 1990-2019.

\input{Table9.tex}

\end{document}

%% file: Table1.tex
\begin{table*}[!ht]
\caption{Hyperparameters tuning: cross-validation results}
\centering
\resizebox{.9\columnwidth}{!}{%
\begin{tabular}{|c|c|c|c|c|c|}
\hline
 \textbf{} & \multirow{3}{*}{\thead{ TA }} & \multirow{3}{*}{\thead{ TS }} & \multirow{3}{*}{\thead{  RBF  }} & \multirow{3}{*}{\thead{TE}} & \multirow{3}{*}{\thead{OnE}}  \\
& & & & & \\
& & & & & \\
\hline
\multirow{3}{*}{\thead{Train: 2015 \\ Val: 2016}} & \multirow{3}{*}{$\lambda=0.1$} & \multirow{2}{*}{$\lambda_1=10$} & \multirow{3}{*}{$\lambda=1$} & \multirow{2}{*}{$\lambda_{diag}=0.01$} & \multirow{3}{*}{$\lambda_{diag}=100$}  \\ 
& & \multirow{2}{*}{$\lambda_2=100$} & & \multirow{2}{*}{$\lambda_{last}=100$} & \\
& & & & & \\
\hline
\multirow{3}{*}{\thead{Train: 2016 \\ Val: 2017}} & \multirow{3}{*}{$\lambda=0.1$} & \multirow{2}{*}{$\lambda_1=10$} & \multirow{3}{*}{$\lambda=10$} & \multirow{2}{*}{$\lambda_{diag}=1$} & \multirow{3}{*}{$\lambda_{diag}=10000$}  \\
& & \multirow{2}{*}{$\lambda_2=1$} & & \multirow{2}{*}{$\lambda_{last}=0.01$} & \\
& & & & & \\
\hline
\multirow{3}{*}{\thead{Train: 2017 \\ Val: 2018}} & \multirow{3}{*}{$\lambda=1$} & \multirow{2}{*}{$\lambda_1=100$} & \multirow{3}{*}{$\lambda=10$} & \multirow{2}{*}{$\lambda_{diag}=0.01$} & \multirow{3}{*}{$\lambda_{diag}=10000$}  \\
& & \multirow{2}{*}{$\lambda_2=10$} & & \multirow{2}{*}{$\lambda_{last}=1$} & \\
& & & & & \\
\hline
\end{tabular}%
}
\label{table:Table1}
\end{table*}

%% file: Table2.tex
\begin{table*}[!htb]
\caption{Forecast performances on 2017, with the percentage variation with respect to Terna results between brackets.}
\centering
\resizebox{\columnwidth}{!}{%
\begin{tabular}{|c|c|c|c|c|c|c|c|}
\hline
 \textbf{} & \multirow{3}{*}{\thead{$\mathbf{MAPE [\%]}$}} & \multirow{3}{*}{\thead{$\mathbf{RMSE [GW]}$}} & \multirow{3}{*}{\thead{$\mathbf{MAE [GW]}$}} & \multirow{3}{*}{\thead{$\mathbf{MAPE_{daily}[\%]}$}} & \multirow{3}{*}{\thead{$\mathbf{RMSE_{daily}[GW]}$}} & \multirow{3}{*}{\thead{$\mathbf{MAE_{daily}[GW]}$}} & \multirow{3}{*}{\thead{$\mathbf{dof}$}} \\
& & & & & & & \\
& & & & & & & \\
\hline
\multirow{3}{*}{\thead{Terna}} & \multirow{3}{*}{$2.16$} & \multirow{3}{*}{$\bf{1.06}$} & \multirow{3}{*}{$0.83$} & \multirow{3}{*}{$\bf{1.22}$} & \multirow{3}{*}{$\bf{0.62}$} & \multirow{3}{*}{$\bf{0.47}$} & \multirow{3}{*}{Unknown} \\ 
& & & & & & & \\
& & & & & & & \\
\hline
\multirow{3}{*}{\thead{OLS}} & \multirow{3}{*}{$2.63\;(22\%)$} & \multirow{3}{*}{$1.41\;(33\%)$} & \multirow{3}{*}{$1.01\;(22\%)$} & \multirow{3}{*}{$1.86\;(52\%)$} & \multirow{3}{*}{$0.95\;(53\%)$} & \multirow{3}{*}{$0.69\;(47\%)$} & \multirow{3}{*}{$9216$} \\ 
& & & & & & & \\
& & & & & & & \\
\hline
\multirow{3}{*}{\thead{TA}} & \multirow{3}{*}{$1.97\;(-9\%)$} & \multirow{3}{*}{$1.07\;(1\%)$} & \multirow{3}{*}{$0.75\;(-10\%)$} & \multirow{3}{*}{$1.42\;(16\%)$} & \multirow{3}{*}{$0.73\;(18\%)$} & \multirow{3}{*}{$0.53\;(13\%)$} & \multirow{3}{*}{$1686.12$} \\ 
& & & & & & & \\
& & & & & & & \\
\hline
\multirow{3}{*}{\thead{TS}} & \multirow{3}{*}{$1.97\;(-9\%)$} & \multirow{3}{*}{$1.07\;(1\%)$} & \multirow{3}{*}{$0.76\;(-8\%)$} & \multirow{3}{*}{$1.41\;(16\%)$} & \multirow{3}{*}{$0.73\;(18\%)$} & \multirow{3}{*}{$0.53\;(13\%)$} & \multirow{3}{*}{$138.38$} \\ 
& & & & & & & \\
& & & & & & & \\
\hline
\multirow{3}{*}{\thead{RBF}} & \multirow{3}{*}{$\bf{1.97\;(-9\%)}$} & \multirow{3}{*}{$1.08\;(2\%)$} & \multirow{3}{*}{$\bf{0.76\;(-8\%)}$} & \multirow{3}{*}{$1.43\;(17\%)$} & \multirow{3}{*}{$0.73\;(18\%)$} & \multirow{3}{*}{$0.54\;(15\%)$} & \multirow{3}{*}{$132.88$} \\
& & & & & & & \\
& & & & & & & \\
\hline
\multirow{3}{*}{\thead{TE}} & \multirow{3}{*}{$2.06\;(-5\%)$} & \multirow{3}{*}{$1.13\;(7\%)$} & \multirow{3}{*}{$0.79\;(-5\%)$} & \multirow{3}{*}{$1.57\;(29\%)$} & \multirow{3}{*}{$0.79\;(27\%)$} & \multirow{3}{*}{$0.59\;(26\%)$} & \multirow{3}{*}{$97.06$} \\
& & & & & & & \\
& & & & & & & \\
\hline
\multirow{3}{*}{\thead{OnE}} & \multirow{3}{*}{$2.64\;(22\%)$} & \multirow{3}{*}{$1.4\;(32\%)$} & \multirow{3}{*}{$0.99\;(19\%)$} & \multirow{3}{*}{$2.04\;(67\%)$} & \multirow{3}{*}{$1.12\;(81\%)$} & \multirow{3}{*}{$0.77\;(64\%)$} & \multirow{3}{*}{$11.12$} \\
& & & & & & & \\
& & & & & & & \\
\hline
\end{tabular}%
}
\label{table:Table2}
\end{table*}

%% file: Table3.tex
\begin{table*}[!htb]
\caption{Forecast performances on 2018, with the percentage variation with respect to Terna results between brackets.}
\centering
\resizebox{\columnwidth}{!}{%
\begin{tabular}{|c|c|c|c|c|c|c|c|}
\hline
 \textbf{} & \multirow{3}{*}{\thead{$\mathbf{MAPE [\%]}$}} & \multirow{3}{*}{\thead{$\mathbf{RMSE [GW]}$}} & \multirow{3}{*}{\thead{$\mathbf{MAE [GW]}$}} & \multirow{3}{*}{\thead{$\mathbf{MAPE_{daily}[\%]}$}} & \multirow{3}{*}{\thead{$\mathbf{RMSE_{daily}[GW]}$}} & \multirow{3}{*}{\thead{$\mathbf{MAE_{daily}[GW]}$}} & \multirow{3}{*}{\thead{$\mathbf{dof}$}} \\
& & & & & & & \\
& & & & & & & \\
\hline
\multirow{3}{*}{\thead{Terna}} & \multirow{3}{*}{$2.41$} & \multirow{3}{*}{$1.14$} & \multirow{3}{*}{$0.9$} & \multirow{3}{*}{$1.65$} & \multirow{3}{*}{$0.76$} & \multirow{3}{*}{$0.62$} & \multirow{3}{*}{Unknown} \\ 
& & & & & & & \\
& & & & & & & \\
\hline
\multirow{3}{*}{\thead{OLS}} & \multirow{3}{*}{$2.74\;(14\%)$} & \multirow{3}{*}{$1.45\;(27\%)$} & \multirow{3}{*}{$1.07\;(19\%)$} & \multirow{3}{*}{$2.0\;(21\%)$} & \multirow{3}{*}{$0.99\;(30\%)$} & \multirow{3}{*}{$0.76\;(23\%)$} & \multirow{3}{*}{$9216$} \\ 
& & & & & & & \\
& & & & & & & \\
\hline
\multirow{3}{*}{\thead{TA}} & \multirow{3}{*}{$1.93\;(-20\%)$} & \multirow{3}{*}{$1.07\;(-6\%)$} & \multirow{3}{*}{$0.75\;(-17\%)$} & \multirow{3}{*}{$1.48\;(-10\%)$} & \multirow{3}{*}{$0.74\;(-3\%)$} & \multirow{3}{*}{$0.55\;(-11\%)$} & \multirow{3}{*}{$1569.78$} \\ 
& & & & & & & \\
& & & & & & & \\
\hline
\multirow{3}{*}{\thead{TS}} & \multirow{3}{*}{$1.93\;(-20\%)$} & \multirow{3}{*}{$1.07\;(-6\%)$} & \multirow{3}{*}{$0.75\;(-17\%)$} & \multirow{3}{*}{$1.49\;(-10\%)$} & \multirow{3}{*}{$0.74\;(-3\%)$} & \multirow{3}{*}{$0.55\;(-11\%)$} & \multirow{3}{*}{$406.03$} \\
& & & & & & & \\
& & & & & & & \\
\hline
\multirow{3}{*}{\thead{RBF}} & \multirow{3}{*}{$1.92\;(-20\%)$} & \multirow{3}{*}{$1.06\;(-7\%)$} & \multirow{3}{*}{$0.74\;(-18\%)$} & \multirow{3}{*}{$1.46\;(-12\%)$} & \multirow{3}{*}{$0.73\;(-4\%)$} & \multirow{3}{*}{$0.55\;(-11\%)$} & \multirow{3}{*}{$80.49$} \\
& & & & & & & \\
& & & & & & & \\
\hline
\multirow{3}{*}{\thead{TE}} & \multirow{3}{*}{$\bf{1.87\;(-22\%)}$} & \multirow{3}{*}{$\bf{1.01\;(-11\%)}$} & \multirow{3}{*}{$\bf{0.72\;(-20\%)}$} & \multirow{3}{*}{$\bf{1.4\;(-15\%)}$} & \multirow{3}{*}{$\bf{0.68\;(-11\%)}$} & \multirow{3}{*}{$\bf{0.52\;(-16\%)}$} & \multirow{3}{*}{$114.5$} \\
& & & & & & & \\
& & & & & & & \\
\hline
\multirow{3}{*}{\thead{OnE}} & \multirow{3}{*}{$2.32\;(-4\%)$} & \multirow{3}{*}{$1.22\;(7\%)$} & \multirow{3}{*}{$0.89\;(-1\%)$} & \multirow{3}{*}{$1.77\;(7\%)$} & \multirow{3}{*}{$0.89\;(17\%)$} & \multirow{3}{*}{$0.67\;(8\%)$} & \multirow{3}{*}{$4.15$} \\
& & & & & & & \\
& & & & & & & \\
\hline
\end{tabular}%
}
\label{table:Table3}
\end{table*}

%% file: Table4.tex
\begin{table*}[!htb]
\caption{Forecast performances on 2019, with the percentage variation with respect to Terna results between brackets.}
\centering
\resizebox{\columnwidth}{!}{%
\begin{tabular}{|c|c|c|c|c|c|c|c|}
\hline
 \textbf{} & \multirow{3}{*}{\thead{$\mathbf{MAPE [\%]}$}} & \multirow{3}{*}{\thead{$\mathbf{RMSE [GW]}$}} & \multirow{3}{*}{\thead{$\mathbf{MAE [GW]}$}} & \multirow{3}{*}{\thead{$\mathbf{MAPE_{daily}[\%]}$}} & \multirow{3}{*}{\thead{$\mathbf{RMSE_{daily}[GW]}$}} & \multirow{3}{*}{\thead{$\mathbf{MAE_{daily}[GW]}$}} & \multirow{3}{*}{\thead{$\mathbf{dof}$}} \\
& & & & & & & \\
& & & & & & & \\
\hline
\multirow{3}{*}{\thead{Terna}} & \multirow{3}{*}{$2.53$} & \multirow{3}{*}{$1.2$} & \multirow{3}{*}{$0.94$} & \multirow{3}{*}{$1.89$} & \multirow{3}{*}{$0.85$} & \multirow{3}{*}{$0.71$} & \multirow{3}{*}{Unknown} \\ 
& & & & & & & \\
& & & & & & & \\
\hline
\multirow{3}{*}{\thead{OLS}} & \multirow{3}{*}{$2.51\;(-1\%)$} & \multirow{3}{*}{$1.33\;(11\%)$} & \multirow{3}{*}{$0.98\;(4\%)$} & \multirow{3}{*}{$1.98\;(5\%)$} & \multirow{3}{*}{$0.92\;(8\%)$} & \multirow{3}{*}{$0.74\;(4\%)$} & \multirow{3}{*}{$9216$} \\ 
& & & & & & & \\
& & & & & & & \\
\hline
\multirow{3}{*}{\thead{TA}} & \multirow{3}{*}{$1.9\;(-25\%)$} & \multirow{3}{*}{$1.02\;(-15\%)$} & \multirow{3}{*}{$0.73\;(-22\%)$} & \multirow{3}{*}{$1.42\;(-25\%)$} & \multirow{3}{*}{$0.72\;(-15\%)$} & \multirow{3}{*}{$0.54\;(-24\%)$} & \multirow{3}{*}{$477.36$} \\ 
& & & & & & & \\
& & & & & & & \\
\hline
\multirow{3}{*}{\thead{TS}} & \multirow{3}{*}{$1.84\;(-27\%)$} & \multirow{3}{*}{$0.99\;(-18\%)$} & \multirow{3}{*}{$0.71\;(-24\%)$} & \multirow{3}{*}{$1.37\;(-28\%)$} & \multirow{3}{*}{$0.68\;(-20\%)$} & \multirow{3}{*}{$0.51\;(-28\%)$} & \multirow{3}{*}{$179.04$} \\
& & & & & & & \\
& & & & & & & \\
\hline
\multirow{3}{*}{\thead{RBF}} & \multirow{3}{*}{$\bf{1.84\;(-27\%)}$} & \multirow{3}{*}{$\bf{0.99\;(-18\%)}$} & \multirow{3}{*}{$\bf{0.71\;(-24\%)}$} & \multirow{3}{*}{$\bf{1.36\;(-28\%)}$} & \multirow{3}{*}{$\bf{0.68\;(-20\%)}$} & \multirow{3}{*}{$\bf{0.51\;(-28\%)}$} & \multirow{3}{*}{$78.42$} \\
& & & & & & & \\
& & & & & & & \\
\hline
\multirow{3}{*}{\thead{TE}} & \multirow{3}{*}{$1.91\;(-25\%)$} & \multirow{3}{*}{$1.04\;(-13\%)$} & \multirow{3}{*}{$0.74\;(-21\%)$} & \multirow{3}{*}{$1.48\;(-22\%)$} & \multirow{3}{*}{$0.73;(-14\%)$} & \multirow{3}{*}{$0.56\;(-21\%)$} & \multirow{3}{*}{$106.37$} \\
& & & & & & & \\
& & & & & & & \\
\hline
\multirow{3}{*}{\thead{OnE}} & \multirow{3}{*}{$2.41\;(-5\%)$} & \multirow{3}{*}{$1.25\;(4\%)$} & \multirow{3}{*}{$0.92\;(-2\%)$} & \multirow{3}{*}{$1.88\;(-1\%)$} & \multirow{3}{*}{$0.97\;(14\%)$} & \multirow{3}{*}{$0.72\;(1\%)$} & \multirow{3}{*}{$3.71$} \\
& & & & & & & \\
& & & & & & & \\
\hline
\end{tabular}%
}
\label{table:Table4}
\end{table*}

%% file: Table5.tex
\begin{table*}[!htb]
\caption{Covariances, biases and Mean Squared Errors of RBF, TE and Terna forecasters, and the corresponding aggregated predictors.}
\centering 
\resizebox{0.4\columnwidth}{!}{%
\begin{tabular}{|c|c|c|c|}
\hline
 \textbf{} & \multirow{3}{*}{\thead{$\mathbf{2017}$}} & \multirow{3}{*}{\thead{$\mathbf{2018}$}} & \multirow{3}{*}{\thead{$\mathbf{2019}$}} \\
& & & \\
& & & \\
\hline
\multirow{3}{*}{\thead{$Cov\left[e_{RBF}e_{Terna}\right]$}} & \multirow{3}{*}{$0.29$} & \multirow{3}{*}{$0.33$} & \multirow{3}{*}{$0.29$} \\ 
& & & \\
& & & \\
\hline
\multirow{3}{*}{\thead{$Cov\left[e_{TE}e_{Terna}\right]$}} & \multirow{3}{*}{$0.36$} & \multirow{3}{*}{$0.36$} & \multirow{3}{*}{$0.31$} \\ 
& & & \\
& & & \\
\hline
\multirow{3}{*}{\thead{$bias_{Terna}$}} & \multirow{3}{*}{$0.16$} & \multirow{3}{*}{$-0.52$} & \multirow{3}{*}{$-0.65$} \\ 
& & & \\
& & & \\
\hline
\multirow{3}{*}{\thead{$bias_{RBF}$}} & \multirow{3}{*}{$-0.03$} & \multirow{3}{*}{$0.02$} & \multirow{3}{*}{$-0.06$} \\ 
& & & \\
& & & \\
\hline
\multirow{3}{*}{\thead{$bias_{TE}$}} & \multirow{3}{*}{$-0.04$} & \multirow{3}{*}{$-0.01$} & \multirow{3}{*}{$-0.09$} \\
& & & \\
& & & \\
\hline
\multirow{3}{*}{\thead{$MSE_{Terna}$}} & \multirow{3}{*}{$1.13$} & \multirow{3}{*}{$1.29$} & \multirow{3}{*}{$1.43$} \\
& & & \\
& & & \\
\hline
\multirow{3}{*}{\thead{$MSE_{RBF}$}} & \multirow{3}{*}{$1.16$} & \multirow{3}{*}{$1.12$} & \multirow{3}{*}{$0.98$} \\
& & & \\
& & & \\
\hline
\multirow{3}{*}{\thead{$MSE_{TE}$}} & \multirow{3}{*}{$1.28$} & \multirow{3}{*}{$1.02$} & \multirow{3}{*}{$1.09$} \\
& & & \\
& & & \\
\hline
\multirow{3}{*}{\thead{$\hat{MSE}_{Avg(RBF)}$}} & \multirow{3}{*}{$0.72$} & \multirow{3}{*}{$0.77$} & \multirow{3}{*}{$0.75$} \\
& & & \\
& & & \\
\hline
\multirow{3}{*}{\thead{$MSE_{Avg(RBF)}$}} & \multirow{3}{*}{$0.71$} & \multirow{3}{*}{$0.77$} & \multirow{3}{*}{$0.77$} \\
& & & \\
& & & \\
\hline
\multirow{3}{*}{\thead{$\hat{MSE}_{Avg(TE)}$}} & \multirow{3}{*}{$0.78$} & \multirow{3}{*}{$0.76$} & \multirow{3}{*}{$0.79$} \\
& & & \\
& & & \\
\hline
\multirow{3}{*}{\thead{$MSE_{Avg(TE)}$}} & \multirow{3}{*}{$0.78$} & \multirow{3}{*}{$0.76$} & \multirow{3}{*}{$0.81$} \\
& & & \\
& & & \\
\hline
\end{tabular}%
}
\label{table:Table5}
\end{table*}

%% file: Table6.tex
\begin{table*}[!htb]
\caption{Aggregated forecast performances on 2017, with the percentage variation with respect to Terna results between brackets.}
\centering
\resizebox{\columnwidth}{!}{%
\begin{tabular}{|c|c|c|c|c|c|c|}
\hline
 \textbf{} & \multirow{3}{*}{\thead{$\mathbf{MAPE [\%]}$}} & \multirow{3}{*}{\thead{$\mathbf{RMSE [GW]}$}} & \multirow{3}{*}{\thead{$\mathbf{MAE [GW]}$}} & \multirow{3}{*}{\thead{$\mathbf{MAPE_{daily}[\%]}$}} & \multirow{3}{*}{\thead{$\mathbf{RMSE_{daily}[GW]}$}} & \multirow{3}{*}{\thead{$\mathbf{MAE_{daily}[GW]}$}} \\
& & & & & & \\
& & & & & & \\
\hline
\multirow{3}{*}{\thead{Terna}} & \multirow{3}{*}{$2.16$} & \multirow{3}{*}{$1.06$} & \multirow{3}{*}{$0.83$} & \multirow{3}{*}{$1.22$} & \multirow{3}{*}{$0.62$} & \multirow{3}{*}{$0.47$} \\ 
& & & & & & \\
& & & & & & \\
\hline
\multirow{3}{*}{\thead{Avg(OLS)}} & \multirow{3}{*}{$1.87\;(-13\%)$} & \multirow{3}{*}{$0.96\;(-9\%)$} & \multirow{3}{*}{$0.72\;(-13\%)$} & \multirow{3}{*}{$1.19\;(-2\%)$} & \multirow{3}{*}{$0.6\;(-3\%)$} & \multirow{3}{*}{$0.45\;(-4\%)$} \\ 
& & & & & & \\
& & & & & & \\
\hline
\multirow{3}{*}{\thead{Avg(TA)}} & \multirow{3}{*}{$1.66\;(-23\%)$} & \multirow{3}{*}{$0.85\;(-20\%)$} & \multirow{3}{*}{$0.63\;(-24\%)$} & \multirow{3}{*}{$1.06\;(-13\%)$} & \multirow{3}{*}{$0.53\;(-15\%)$} & \multirow{3}{*}{$0.4\;(-15\%)$} \\ 
& & & & & & \\
& & & & & & \\
\hline
\multirow{3}{*}{\thead{Avg(TS)}} & \multirow{3}{*}{$1.66\;(-23\%)$} & \multirow{3}{*}{$0.85\;(-20\%)$} & \multirow{3}{*}{$0.63\;(-24\%)$} & \multirow{3}{*}{$1.04\;(-15\%)$} & \multirow{3}{*}{$0.52\;(-16\%)$} & \multirow{3}{*}{$0.39\;(-17\%)$} \\
& & & & & & \\
& & & & & & \\
\hline
\multirow{3}{*}{\thead{Avg(RBF)}} & \multirow{3}{*}{$\bf{1.66\;(-23\%)}$} & \multirow{3}{*}{$\bf{0.85\;(-20\%)}$} & \multirow{3}{*}{$\bf{0.63\;(-24\%)}$} & \multirow{3}{*}{$\bf{1.04\;(-15\%)}$} & \multirow{3}{*}{$\bf{0.52\;(-16\%)}$} & \multirow{3}{*}{$\bf{0.39\;(-17\%)}$} \\
& & & & & & \\
& & & & & & \\
\hline
\multirow{3}{*}{\thead{Avg(TE)}} & \multirow{3}{*}{$1.72\;(-20\%)$} & \multirow{3}{*}{$0.88\;(-17\%)$} & \multirow{3}{*}{$0.66\;(-20\%)$} & \multirow{3}{*}{$1.13\;(-7\%)$} & \multirow{3}{*}{$0.57\;(-8\%)$} & \multirow{3}{*}{$0.42\;(-11\%)$} \\
& & & & & & \\
& & & & & & \\
\hline
\multirow{3}{*}{\thead{Avg(OnE)}} & \multirow{3}{*}{$2.04\;(-6\%)$} & \multirow{3}{*}{$1.03\;(-3\%)$} & \multirow{3}{*}{$0.77\;(-7\%)$} & \multirow{3}{*}{$1.44\;(18\%)$} & \multirow{3}{*}{$0.76\;(23\%)$} & \multirow{3}{*}{$0.55\;(17\%)$} \\
& & & & & & \\
& & & & & & \\
\hline
\end{tabular}%
}
\label{table:Table6}
\end{table*}

%% file: Table7.tex
\begin{table*}[!htb]
\caption{Aggregated forecast performances on 2018, with the percentage variation with respect to Terna results between brackets.}
\centering
\resizebox{\columnwidth}{!}{%
\begin{tabular}{|c|c|c|c|c|c|c|}
\hline
 \textbf{} & \multirow{3}{*}{\thead{$\mathbf{MAPE [\%]}$}} & \multirow{3}{*}{\thead{$\mathbf{RMSE [GW]}$}} & \multirow{3}{*}{\thead{$\mathbf{MAE [GW]}$}} & \multirow{3}{*}{\thead{$\mathbf{MAPE_{daily}[\%]}$}} & \multirow{3}{*}{\thead{$\mathbf{RMSE_{daily}[GW]}$}} & \multirow{3}{*}{\thead{$\mathbf{MAE_{daily}[GW]}$}} \\
& & & & & & \\
& & & & & & \\
\hline
\multirow{3}{*}{\thead{Terna}} & \multirow{3}{*}{$2.41$} & \multirow{3}{*}{$1.14$} & \multirow{3}{*}{$0.9$} & \multirow{3}{*}{$1.65$} & \multirow{3}{*}{$0.76$} & \multirow{3}{*}{$0.62$} \\ 
& & & & & & \\
& & & & & & \\
\hline
\multirow{3}{*}{\thead{Avg(OLS)}} & \multirow{3}{*}{$1.94\;(-20\%)$} & \multirow{3}{*}{$1.01\;(-11\%)$} & \multirow{3}{*}{$0.74\;(-18\%)$} & \multirow{3}{*}{$1.32\;(-20\%)$} & \multirow{3}{*}{$0.66\;(-13\%)$} & \multirow{3}{*}{$0.49\;(-21\%)$} \\ 
& & & & & & \\
& & & & & & \\
\hline
\multirow{3}{*}{\thead{Avg(TA)}} & \multirow{3}{*}{$1.71\;(-29\%)$} & \multirow{3}{*}{$0.88\;(-23\%)$} & \multirow{3}{*}{$0.65\;(-28\%)$} & \multirow{3}{*}{$1.16\;(-30\%)$} & \multirow{3}{*}{$0.58\;(-24\%)$} & \multirow{3}{*}{$0.43\;(-31\%)$} \\ 
& & & & & & \\
& & & & & & \\
\hline
\multirow{3}{*}{\thead{Avg(TS)}} & \multirow{3}{*}{$\bf{1.7\;(-29\%)}$} & \multirow{3}{*}{$0.87\;(-24\%)$} & \multirow{3}{*}{$0.65\;(-28\%)$} & \multirow{3}{*}{$\bf{1.15\;(-30\%)}$} & \multirow{3}{*}{$0.57\;(-25\%)$} & \multirow{3}{*}{$\bf{0.43\;(-31\%)}$} \\
& & & & & & \\
& & & & & & \\
\hline
\multirow{3}{*}{\thead{Avg(RBF)}} & \multirow{3}{*}{$1.71\;(-29\%)$} & \multirow{3}{*}{$0.88\;(-23\%)$} & \multirow{3}{*}{$0.65\;(-28\%)$} & \multirow{3}{*}{$1.16\;(-30\%)$} & \multirow{3}{*}{$0.58\;(-24\%)$} & \multirow{3}{*}{$0.43\;(-31\%)$} \\
& & & & & & \\
& & & & & & \\
\hline
\multirow{3}{*}{\thead{Avg(TE)}} & \multirow{3}{*}{$1.72\;(-29\%)$} & \multirow{3}{*}{$\bf{0.87\;(-24\%)}$} & \multirow{3}{*}{$\bf{0.65\;(-28\%)}$} & \multirow{3}{*}{$1.18\;(-28\%)$} & \multirow{3}{*}{$\bf{0.57\;(-25\%)}$} & \multirow{3}{*}{$0.44\;(-29\%)$} \\
& & & & & & \\
& & & & & & \\
\hline
\multirow{3}{*}{\thead{Avg(OnE)}} & \multirow{3}{*}{$1.9\;(-21\%)$} & \multirow{3}{*}{$0.96\;(-16\%)$} & \multirow{3}{*}{$0.72\;(-20\%)$} & \multirow{3}{*}{$1.34\;(-19\%)$} & \multirow{3}{*}{$0.67\;(-12\%)$} & \multirow{3}{*}{$0.5\;(-19\%)$} \\
& & & & & & \\
& & & & & & \\
\hline
\end{tabular}%
}
\label{table:Table7}
\end{table*}

%% file: Table8.tex
\begin{table*}[!htb]
\caption{Aggregated forecast performances on 2019, with the percentage variation with respect to Terna results between brackets.}
\centering
\resizebox{\columnwidth}{!}{%
\begin{tabular}{|c|c|c|c|c|c|c|}
\hline
 \textbf{} & \multirow{3}{*}{\thead{$\mathbf{MAPE [\%]}$}} & \multirow{3}{*}{\thead{$\mathbf{RMSE [GW]}$}} & \multirow{3}{*}{\thead{$\mathbf{MAE [GW]}$}} & \multirow{3}{*}{\thead{$\mathbf{MAPE_{daily}[\%]}$}} & \multirow{3}{*}{\thead{$\mathbf{RMSE_{daily}[GW]}$}} & \multirow{3}{*}{\thead{$\mathbf{MAE_{daily}[GW]}$}} \\
& & & & & & \\
& & & & & & \\
\hline
\multirow{3}{*}{\thead{Terna}} & \multirow{3}{*}{$2.53$} & \multirow{3}{*}{$1.2$} & \multirow{3}{*}{$0.94$} & \multirow{3}{*}{$1.89$} & \multirow{3}{*}{$0.85$} & \multirow{3}{*}{$0.71$} \\ 
& & & & & & \\
& & & & & & \\
\hline
\multirow{3}{*}{\thead{Avg(OLS)}} & \multirow{3}{*}{$1.95\;(-23\%)$} & \multirow{3}{*}{$0.99\;(-18\%)$} & \multirow{3}{*}{$0.74\;(-21\%)$} & \multirow{3}{*}{$1.45\;(-23\%)$} & \multirow{3}{*}{$0.69\;(-19\%)$} & \multirow{3}{*}{$0.54\;(-24\%)$} \\ 
& & & & & & \\
& & & & & & \\
\hline
\multirow{3}{*}{\thead{Avg(TA)}} & \multirow{3}{*}{$1.76\;(-30\%)$} & \multirow{3}{*}{$0.89\;(-26\%)$} & \multirow{3}{*}{$0.67\;(-29\%)$} & \multirow{3}{*}{$1.28\;(-32\%)$} & \multirow{3}{*}{$0.63\;(-26\%)$} & \multirow{3}{*}{$0.48\;(-32\%)$} \\ 
& & & & & & \\
& & & & & & \\
\hline
\multirow{3}{*}{\thead{Avg(TS)}} & \multirow{3}{*}{$1.73\;(-32\%)$} & \multirow{3}{*}{$0.88\;(-27\%)$} & \multirow{3}{*}{$0.65\;(-31\%)$} & \multirow{3}{*}{$1.23\;(-35\%)$} & \multirow{3}{*}{$0.6\;(-29\%)$} & \multirow{3}{*}{$0.46\;(-35\%)$} \\
& & & & & & \\
& & & & & & \\
\hline
\multirow{3}{*}{\thead{Avg(RBF)}} & \multirow{3}{*}{$\bf{1.73\;(-32\%)}$} & \multirow{3}{*}{$\bf{0.88\;(-27\%)}$} & \multirow{3}{*}{$\bf{0.65\;(-31\%)}$} & \multirow{3}{*}{$\bf{1.23\;(-35\%)}$} & \multirow{3}{*}{$\bf{0.6\;(-29\%)}$} & \multirow{3}{*}{$\bf{0.46\;(-35\%)}$} \\
& & & & & & \\
& & & & & & \\
\hline
\multirow{3}{*}{\thead{Avg(TE)}} & \multirow{3}{*}{$1.78\;(-30\%)$} & \multirow{3}{*}{$0.9\;(-25\%)$} & \multirow{3}{*}{$0.67\;(-29\%)$} & \multirow{3}{*}{$1.3\;(-31\%)$} & \multirow{3}{*}{$0.64\;(-25\%)$} & \multirow{3}{*}{$0.49\;(-31\%)$} \\
& & & & & & \\
& & & & & & \\
\hline
\multirow{3}{*}{\thead{Avg(OnE)}} & \multirow{3}{*}{$1.98\;(-22\%)$} & \multirow{3}{*}{$1.0\;(-17\%)$} & \multirow{3}{*}{$0.75\;(-20\%)$} & \multirow{3}{*}{$1.51\;(-20\%)$} & \multirow{3}{*}{$0.74\;(-13\%)$} & \multirow{3}{*}{$0.57\;(-20\%)$} \\
& & & & & & \\
& & & & & & \\
\hline
\end{tabular}%
}
\label{table:Table8}
\end{table*}

%% file: Table9.tex
\begin{table}[ht]
\caption{Easter holiday dates: years 1990-2019.}
\centering 
\resizebox{0.35\columnwidth}{!}{%
\begin{tabular}{|c|c|}
\hline
 \textbf{Year} & \textbf{Date} \\
 \hline
 $1990$ & April 12 - April 16 \\
 \hline
 $1991$ & March 28 - April 1 \\
 \hline
 $1992$ & April 16 - April 20 \\
 \hline
 $1993$ & April 8 - April 12 \\
 \hline
 $1994$ & March 31 - April 4 \\
 \hline
 $1995$ & April 13 - April 17 \\
 \hline
 $1996$ & April 4 - April 8 \\
 \hline
 $1997$ & March 27 - March 31 \\
 \hline
 $1998$ & April 9 - April 13 \\
 \hline
 $1999$ & April 1 - April 5 \\
 \hline
 $2000$ & April 20 - April 24 \\
 \hline
 $2001$ & April 12 - April 16 \\
 \hline
 $2002$ & March 28 - April 1 \\
 \hline
 $2003$ & April 17 - April 21 \\
 \hline
 $2004$ & April 8 - April 12 \\
 \hline
 $2005$ & March 24 - March 28 \\
 \hline
 $2006$ & April 13 - April 17 \\
 \hline
 $2007$ & April 5 - April 9 \\
 \hline
 $2008$ & March 20 - March 24 \\
 \hline
 $2009$ & April 9 - April 13 \\
 \hline
 $2010$ & April 1 - April 5 \\
 \hline
 $2011$ & April 21 - April 25 \\
 \hline
 $2012$ & April 5 - April 9 \\
 \hline
 $2013$ & March 28 - April 1 \\
 \hline
 $2014$ & April 17 - April 21 \\
 \hline
 $2015$ & April 2 - April 6 \\
 \hline
 $2016$ & March 24 - March 28 \\
 \hline
 $2017$ & April 13 - April 17 \\
 \hline
 $2018$ & March 29 - April 2 \\
 \hline
 $2019$ & April 18 - April 22 \\
 \hline
\end{tabular}%
}
\label{table:Table9}
\end{table}

%% file: regularization_short_term.bbl
\begin{thebibliography}{10}
\expandafter\ifx\csname url\endcsname\relax
  \def\url#1{\texttt{#1}}\fi
\expandafter\ifx\csname urlprefix\endcsname\relax\def\urlprefix{URL }\fi
\expandafter\ifx\csname href\endcsname\relax
  \def\href#1#2{#2} \def\path#1{#1}\fi

\bibitem{christiaanse1971short}
W.~Christiaanse, Short-term load forecasting using general exponential
  smoothing, IEEE Transactions on Power Apparatus and Systems (1971)
  900--911\href {http://dx.doi.org/10.1109/TPAS.1971.293123}
  {\path{doi:10.1109/TPAS.1971.293123}}.

\bibitem{nbamalu1993autoregressive}
G.~Mbamalu, M.~El-Hawary, Load forecasting via suboptimal seasonal
  autoregressive models and iteratively reweighted least squares estimation,
  IEEE Transactions on Power Systems 8~(1) (1993) 343--348.
\newblock \href {http://dx.doi.org/10.1109/59.221222}
  {\path{doi:10.1109/59.221222}}.

\bibitem{chen1995arma}
J.-F. Chen, W.-M. Wang, C.-M. Huang, Analysis of an adaptive time-series
  autoregressive moving-average (arma) model for short-term load forecasting,
  Electric Power Systems Research 34~(3) (1995) 187--196.
\newblock \href
  {http://dx.doi.org/https://doi.org/10.1016/0378-7796(95)00977-1}
  {\path{doi:https://doi.org/10.1016/0378-7796(95)00977-1}}.

\bibitem{huang1997thresholdautoregressive}
S.~Huang, Short-term load forecasting using threshold autoregressive models,
  IEE Proceedings-Generation, Transmission and Distribution 144~(5) (1997)
  477--481.
\newblock \href {http://dx.doi.org/10.1049/ip-gtd:19971144}
  {\path{doi:10.1049/ip-gtd:19971144}}.

\bibitem{soares2005seasonalautoregressive}
L.~J. Soares, M.~C. Medeiros,
  \href{https://ideas.repec.org/p/rio/texdis/495.html}{{Modelling and
  forecasting short-term electricity load: a two step methodology}}, Textos
  para discussão 495, Department of Economics PUC-Rio (Brazil) (Feb. 2005).
\newline\urlprefix\url{https://ideas.repec.org/p/rio/texdis/495.html}

\bibitem{yang1995armax}
{Hong-Tzer Yang}, {Chao-Ming Huang}, {Ching-Lien Huang}, Identification of
  armax model for short term load forecasting: an evolutionary programming
  approach, in: Proceedings of Power Industry Computer Applications Conference,
  1995, pp. 325--330.

\bibitem{charytoniuk1998nonparametricregression}
W.~{Charytoniuk}, M.~S. {Chen}, P.~{Van Olinda}, Nonparametric regression based
  short-term load forecasting, IEEE Transactions on Power Systems 13~(3) (1998)
  725--730.

\bibitem{hyndman2012semiparametric}
S.~Fan, R.~J. Hyndman, Short-term load forecasting based on a semi-parametric
  additive model, IEEE Transactions on Power Systems 27.

\bibitem{dordonnat2016semiparametric}
V.~Dordonnat, A.~Pichavant, A.~Pierrot,
  \href{https://ideas.repec.org/a/eee/intfor/v32y2016i3p1005-1011.html}{{GEFCom2014
  probabilistic electric load forecasting using time series and semi-parametric
  regression models}}, International Journal of Forecasting 32~(3) (2016)
  1005--1011.
\newblock \href {http://dx.doi.org/10.1016/j.ijforecast.2015}
  {\path{doi:10.1016/j.ijforecast.2015}}.
\newline\urlprefix\url{https://ideas.repec.org/a/eee/intfor/v32y2016i3p1005-1011.html}

\bibitem{moghram1989fiveshorttermforecastingtechniques}
I.~{Moghram}, S.~{Rahman}, Analysis and evaluation of five short-term load
  forecasting techniques, IEEE Transactions on Power Systems 4~(4) (1989)
  1484--1491.

\bibitem{alhamadi2004kalmanfilter}
H.~Al-Hamadi, S.~Soliman, Short-term electric load forecasting based on kalman
  filtering algorithm with moving window weather and load model, Electric Power
  Systems Research - ELEC POWER SYST RES 68 (2004) 47--59.
\newblock \href {http://dx.doi.org/10.1016/S0378-7796(03)00150-0}
  {\path{doi:10.1016/S0378-7796(03)00150-0}}.

\bibitem{takeda2016kalmanfilter}
H.~Takeda, Y.~Tamura, S.~Sato, Using the ensemble kalman filter for electricity
  load forecasting and analysis, Energy 104 (2016) 184--198.
\newblock \href {http://dx.doi.org/10.1016/j.energy.2016.03.070}
  {\path{doi:10.1016/j.energy.2016.03.070}}.

\bibitem{kuo2018artificialneuralnetwork}
P.-H. Kuo, C.-J. Huang, A high precision artificial neural networks model for
  short-term energy load forecasting, Energies 11 (2018) 213.
\newblock \href {http://dx.doi.org/10.3390/en11010213}
  {\path{doi:10.3390/en11010213}}.

\bibitem{ryu2017deepneuralnetwork}
S.~Ryu, J.~Noh, H.~Kim, Deep neural network based demand side short term load
  forecasting, Energies 10 (2016) 3.
\newblock \href {http://dx.doi.org/10.3390/en10010003}
  {\path{doi:10.3390/en10010003}}.

\bibitem{kong2017lstmneuralnetwork}
W.~Kong, Z.~Dong, Y.~Jia, D.~Hill, Y.~Xu, Y.~Zhang, Short-term residential load
  forecasting based on lstm recurrent neural network, IEEE Transactions on
  Smart Grid PP (2017) 1--1.
\newblock \href {http://dx.doi.org/10.1109/TSG.2017.2753802}
  {\path{doi:10.1109/TSG.2017.2753802}}.

\bibitem{yun2008rbf}
Z.~Yun, Z.~Quan, S.~Caixin, L.~Shaolan, L.~Yuming, S.~Yang, Rbf neural network
  and anfis-based short-term load forecasting approach in real-time price
  environment, IEEE Transactions on power systems 23~(3) (2008) 853--858.

\bibitem{cecati2015rbfshortterm}
C.~Cecati, J.~Kolbusz, P.~R{\'o}{\.z}ycki, P.~Siano, B.~M. Wilamowski, A novel
  rbf training algorithm for short-term electric load forecasting and
  comparative studies, IEEE Transactions on industrial Electronics 62~(10)
  (2015) 6519--6529.

\bibitem{jalali2021novel}
S.~M.~J. Jalali, S.~Ahmadian, A.~Khosravi, M.~Shafie-khah, S.~Nahavandi, J.~P.
  Catalao, A novel evolutionary-based deep convolutional neural network model
  for intelligent load forecasting, IEEE Transactions on Industrial
  Informatics\href {http://dx.doi.org/10.1109/TII.2021.3065718}
  {\path{doi:10.1109/TII.2021.3065718}}.

\bibitem{hassan2016fuzzy}
S.~Hassan, A.~Khosravi, J.~Jaafar, M.~Khanesar, A systematic design of interval
  type-2 fuzzy logic system using extreme learning machine for electricity load
  demand forecasting, International Journal of Electrical Power \& Energy
  Systems 82 (2016) 1--10.
\newblock \href {http://dx.doi.org/10.1016/j.ijepes.2016.03.001}
  {\path{doi:10.1016/j.ijepes.2016.03.001}}.

\bibitem{khosravi2013load}
A.~Khosravi, S.~Nahavandi, Load forecasting using interval type-2 fuzzy logic
  systems: Optimal type reduction, IEEE Transactions on Industrial Informatics
  10~(2) (2013) 1055--1063.
\newblock \href {http://dx.doi.org/10.1109/TII.2013.2285650}
  {\path{doi:10.1109/TII.2013.2285650}}.

\bibitem{zhang2017svm}
X.~Zhang, J.~Wang, K.~Zhang, Short-term electric load forecasting based on
  singular spectrum analysis and support vector machine optimized by cuckoo
  search algorithm, Electric Power Systems Research 146 (2017) 270--285.
\newblock \href {http://dx.doi.org/10.1016/j.epsr.2017.01.035}
  {\path{doi:10.1016/j.epsr.2017.01.035}}.

\bibitem{jiang2018svr}
H.~Jiang, Y.~Zhang, E.~Muljadi, J.~Zhang, W.~Gao, A short-term and
  high-resolution distribution system load forecasting approach using support
  vector regression with hybrid parameters optimization, IEEE Transactions on
  Smart Grid PP (2016) 1949--3053.
\newblock \href {http://dx.doi.org/10.1109/TSG.2016.2628061}
  {\path{doi:10.1109/TSG.2016.2628061}}.

\bibitem{hagan1987thetimeseriesapproachtoshorttermloadforecasting}
M.~T. {Hagan}, S.~M. {Behr}, The time series approach to short term load
  forecasting, IEEE Transactions on Power Systems 2~(3) (1987) 785--791.
\newblock \href {http://dx.doi.org/10.1109/TPWRS.1987.4335210}
  {\path{doi:10.1109/TPWRS.1987.4335210}}.

\bibitem{sood2010autocorrelation}
R.~Sood, I.~Koprinska, V.~G. Agelidis, Electricity load forecasting based on
  autocorrelation analysis, in: The 2010 International Joint Conference on
  Neural Networks (IJCNN), IEEE, 2010, pp. 1--8.
\newblock \href {http://dx.doi.org/10.1109/IJCNN.2010.5596877}
  {\path{doi:10.1109/IJCNN.2010.5596877}}.

\bibitem{yadav2010autocorrelationsom}
V.~Yadav, D.~Srinivasan, Autocorrelation based weighing strategy for short-term
  load forecasting with the self-organizing map, 2010 The 2nd International
  Conference on Computer and Automation Engineering, ICCAE 2010 1.
\newblock \href {http://dx.doi.org/10.1109/ICCAE.2010.5451972}
  {\path{doi:10.1109/ICCAE.2010.5451972}}.

\bibitem{koprinska2015autocorrelation}
I.~Koprinska, M.~Rana, V.~Agelidis, Correlation and instance based feature
  selection for electricity load forecasting, Knowledge-Based Systems 82.
\newblock \href {http://dx.doi.org/10.1016/j.knosys.2015.02.017}
  {\path{doi:10.1016/j.knosys.2015.02.017}}.

\bibitem{kilian2017var}
L.~Kilian, H.~L{\"u}tkepohl, Structural vector autoregressive analysis,
  Cambridge University Press, 2017.

\bibitem{lutkepohl2005multipletimeseries}
H.~L{\"u}tkepohl, New introduction to multiple time series analysis, Springer
  Science \& Business Media, 2005.

\bibitem{wang2020tensor}
D.~Wang, Y.~Zheng, H.~Lian, G.~Li, High-dimensional vector autoregressive time
  series modeling via tensor decomposition, Journal of the American Statistical
  Association (2021) 1--19.

\bibitem{krampe2021varsparsity}
J.~Krampe, E.~Paparoditis, Sparsity concepts and estimation procedures for
  high-dimensional vector autoregressive models, Journal of Time Series
  Analysis.

\bibitem{ballarin2021varridge}
G.~Ballarin, Ridge regularized estimation of var models for inference and sieve
  approximation, arXiv preprint arXiv:2105.00860.

\bibitem{ternaWebsite}
\url{https://www.terna.it/it/sistema-elettrico/transparency-report/total-load},
  accessed: 2020-04-30.

\bibitem{nowicka2002modeling}
J.~Nowicka-Zagrajek, R.~Weron, Modeling electricity loads in california: Arma
  models with hyperbolic noise, Signal Processing 82~(12) (2002) 1903--1915.
\newblock \href {http://dx.doi.org/10.1016/S0165-1684(02)00318-3}
  {\path{doi:10.1016/S0165-1684(02)00318-3}}.

\bibitem{incremona2020lassofft}
A.~Incremona, G.~De~Nicolao, Spectral characterization of the multi-seasonal
  component of the italian electric load: a lasso-fft approach, IEEE Control
  Systems Letters 4~(1) (2019) 187--192.
\newblock \href {http://dx.doi.org/10.1109/LCSYS.2019.2922192}
  {\path{doi:10.1109/LCSYS.2019.2922192}}.

\bibitem{hong1998weighted}
C.~Hong, L.~Jianwei, A weighted multi-model short-term load forecasting system,
  in: POWERCON'98. 1998 International Conference on Power System Technology.
  Proceedings (Cat. No. 98EX151), Vol.~1, IEEE, 1998, pp. 557--561.
\newblock \href {http://dx.doi.org/10.1109/ICPST.1998.729026}
  {\path{doi:10.1109/ICPST.1998.729026}}.

\bibitem{ahmia2015multimodel}
O.~Ahmia, N.~Farah, Multi-model approach for electrical load forecasting, in:
  2015 SAI Intelligent Systems Conference (IntelliSys), IEEE, 2015, pp. 87--92.
\newblock \href {http://dx.doi.org/10.1109/IntelliSys.2015.7361089}
  {\path{doi:10.1109/IntelliSys.2015.7361089}}.

\bibitem{gruber2017shrinkage}
M.~H. Gruber, Improving efficiency by shrinkage: the James-Stein and ridge
  regression estimators, Routledge, 2017.
\newblock \href {http://dx.doi.org/https://doi.org/10.1201/9780203751220}
  {\path{doi:https://doi.org/10.1201/9780203751220}}.

\bibitem{saleh2019ridge}
A.~M.~E. Saleh, M.~Arashi, B.~G. Kibria, Theory of ridge regression estimation
  with applications, Vol. 285, John Wiley \& Sons, 2019.

\bibitem{ranaweera1995rbf}
D.~Ranaweera, N.~Hubele, A.~Papalexopoulos, Application of radial basis
  function neural network model for short-term load forecasting, Generation,
  Transmission and Distribution, IEE Proceedings- 142 (1995) 45 -- 50.
\newblock \href {http://dx.doi.org/10.1049/ip-gtd:19951602}
  {\path{doi:10.1049/ip-gtd:19951602}}.

\bibitem{konishi2004rbf}
S.~Konishi, Bayesian information criteria and smoothing parameter selection in
  radial basis function networks, Biometrika 91 (2004) 27--43.
\newblock \href {http://dx.doi.org/10.1093/biomet/91.1.27}
  {\path{doi:10.1093/biomet/91.1.27}}.

\bibitem{webb2004ensemblelearning}
G.~I. {Webb}, Z.~{Zheng}, Multistrategy ensemble learning: reducing error by
  combining ensemble learning techniques, IEEE Transactions on Knowledge and
  Data Engineering 16~(8) (2004) 980--991.
\newblock \href {http://dx.doi.org/10.1109/TKDE.2004.29}
  {\path{doi:10.1109/TKDE.2004.29}}.

\bibitem{nowotarski2015sisterforecasts}
J.~Nowotarski, B.~Liu, R.~Weron, T.~Hong, Improving short term load forecast
  accuracy via combining sister forecasts, Energy 98 (2016) 40--49.
\newblock \href {http://dx.doi.org/10.1016/j.energy.2015.12.142}
  {\path{doi:10.1016/j.energy.2015.12.142}}.

\bibitem{wallis2011combiningforecasts}
K.~F. Wallis, Combining forecasts – forty years later, Applied Financial
  Economics 21~(1-2) (2011) 33--41.
\newblock \href {http://dx.doi.org/10.1080/09603107.2011.523179}
  {\path{doi:10.1080/09603107.2011.523179}}.

\bibitem{ranjan2010combiningprobabilityforecasts}
R.~Ranjan, T.~Gneiting, Combining probability forecasts, Journal of the Royal
  Statistical Society Series B 72 (2010) 71--91.
\newblock \href {http://dx.doi.org/10.2307/40541575}
  {\path{doi:10.2307/40541575}}.

\bibitem{devaine2013forecasting}
M.~Devaine, P.~Gaillard, Y.~Goude, G.~Stoltz, Forecasting electricity
  consumption by aggregating specialized experts, Machine Learning 90~(2)
  (2013) 231--260.
\newblock \href {http://dx.doi.org/10.1007/s10994-012-5314-7}
  {\path{doi:10.1007/s10994-012-5314-7}}.

\end{thebibliography}
